\documentclass[10pt,a4paper]{article}
\usepackage{parskip}

\usepackage[T1]{fontenc}       
\usepackage{amssymb}
\usepackage{amsmath}
\usepackage{braket}
\usepackage{graphicx,caption}
\usepackage{multicol}
\usepackage{scrextend}
\usepackage{lineno}
\usepackage{lettrine}
\usepackage{pdfpages}
\usepackage[normalem]{ulem}
\renewcommand{\thesection}{\Roman{section}} 
\renewcommand{\thesubsection}{\thesection.\Alph{subsection}}

\usepackage{titlesec}
\titleformat{\section}
  {\normalfont\fontsize{10}{15}\bfseries}{\thesection}{1em}{}
\titleformat{\subsection}
  {\normalfont\fontsize{10}{15}\bfseries}{\thesubsection}{1em}{}

\usepackage[utf8]{inputenc}
\usepackage{amsfonts}

\newcounter{extfigure} 

\graphicspath{{./Figures/}}

\usepackage[sort&compress,square]{natbib}
\makeatletter
\newcommand{\overleftrightsmallarrow}{\mathpalette{\overarrowsmall@\leftrightarrowfill@}}
\newcommand{\overrightsmallarrow}{\mathpalette{\overarrowsmall@\rightarrowfill@}}
\newcommand{\overleftsmallarrow}{\mathpalette{\overarrowsmall@\leftarrowfill@}}
\newcommand{\overarrowsmall@}[3]{%
	\vbox{%
		\ialign{%
			##\crcr
			#1{\smaller@style{#2}}\crcr
			\noalign{\nointerlineskip}%
			$\m@th\hfil#2#3\hfil$\crcr
		}%
	}%
}
\def\smaller@style#1{%
	\ifx#1\displaystyle\scriptstyle\else
	\ifx#1\textstyle\scriptstyle\else
	\scriptscriptstyle
	\fi
	\fi
}
\makeatother
\newcommand{\tensor}[1]{\overleftrightsmallarrow{#1}}

\begin{document}
	\begin{center}
		
		{\large{\bf
			A singlet-triplet hole-spin qubit in MOS silicon.
			}}
			
			
			\vskip0.5\baselineskip
			
			{\bf
				S. D. Liles$^{1,\dagger}$, D. J. Halverson$^{1}$, Z. Wang$^{1}$, A. Shamim$^{1}$, R. S. Eggli$^{2}$, I. K. Jin$^{1,3}$, J. Hillier$^{1}$, K. Kumar$^{1}$, I. Vorreiter$^{1}$, M. Rendell$^{1}$, J. H. Huang$^{4,5}$, C. C. Escott$^{4,5}$, F. E. Hudson$^{4,5}$, W. H. Lim$^{4,5}$, D. Culcer$^{1}$, A. S. Dzurak$^{4,5}$, A. R. Hamilton$^{1}$  
			}
			
			\vskip0.5\baselineskip
			
			{\it
				$^{1}$School of Physics, University of New South Wales, Sydney NSW 2052, Australia.\\
				$^{2}$Department of Physics, University of Basel, Klingelbergstrasse 82, CH-4056 Basel, Switzerland.\\
				$^{3}$ RIKEN, 2-1, Hirosawa, Wako-shi, Saitama 351-0198, Japan.\\
				$^{4}$School of Electrical Engineering and Telecommunications,\\
				University of New South Wales, Sydney NSW 2052, Australia.\\ 
    			$^{5}$Diraq, Sydney NSW, Australia.\\ 
			}
			
			%
			
			\let\thefootnote\relax\footnote{$\dagger$ Corresponding author - s.liles@unsw.edu.au\\}
\\
\textbf{Abstract}\vspace{1ex}\\ \parbox{0.8\textwidth}{
Holes in silicon quantum dots are promising for spin qubit applications due to the strong intrinsic spin-orbit coupling. The spin-orbit coupling produces complex hole-spin dynamics, providing opportunities to further optimize spin qubits. Here, we demonstrate a singlet-triplet qubit using hole states in a planar metal-oxide-semiconductor double quantum dot. We demonstrate rapid qubit control with singlet-triplet oscillations up to 400~MHz. The qubit exhibits promising coherence, with a maximum dephasing time of 600~ns, which is enhanced to 1.3~$\mu$s using refocusing techniques. We investigate the magnetic field anisotropy of the eigenstates, and determine a magnetic field orientation to improve the qubit initialisation fidelity. These results present a step forward for spin qubit technology, by implementing a high quality singlet-triplet hole-spin qubit in planar architecture suitable for scaling up to 2D arrays of coupled qubits.\\
} 
\end{center}	
			
\begin{multicols}{2}
\subsection*{Introduction}
Spin qubits in group IV materials are promising for semiconductor-based quantum computation applications\cite{loss1998quantum,zwanenburg2013silicon,veldhorst2017silicon}. The most straightforward spin qubit is the single-spin qubit (Loss-DiVincenzo qubit\citep{loss1998quantum}), which encodes information using the $\ket{\uparrow}$ and $\ket{\downarrow}$ spin states. An alternative is the singlet-triplet qubit, which uses the singlet (S = ($\ket{\uparrow\downarrow}-\ket{\downarrow\uparrow})/ \sqrt{2}$) and unpolarised-triplet ($T_0$ = ($\ket{\uparrow\downarrow}+\ket{\downarrow\uparrow})/ \sqrt{2}$) states of two exchange-coupled spins\cite{levy2002universal,petta2005coherent,maune2012coherent,jock2018silicon}. While using two spins rather than one increases the fabrication footprint and the complexity of the eigenstates, singlet-triplet qubits offer advantages over single-spin qubits\cite{burkard2023semiconductor}. Singlet-triplet qubits can be operated at very low magnetic fields (<5 mT), which enables compatibility with magnetic-sensitive components such as superconducting resonators\cite{harvey2018coupling,burkard2020superconductor,bottcher2022parametric}. Additionally, singlet-triplet qubits can be controlled using lower frequency control pulses, with spectral components generally not exceeding 100~MHz. This reduces the cost and complexity of control hardware compared with single-spin qubits, which typically require GHz phase-controlled tones. Removing these GHz control tones has advantages since the power they dissipate can degrade qubit quality\cite{freer2017single,watson2018programmable,philips2022universal,undseth2023hotter}. Further, developing singlet-triplet qubits provides technological advances, since singlet-triplet systems form the building blocks for novel devices including exchange-only\cite{divincenzo2000universal} and resonant-exchange qubits\cite{medford2013quantum,taylor2013electrically}. 

Hole-spins in Group IV materials offer significant opportunities for use as fast coherent spin qubits \cite{maurand2016cmos,camenzind2022hole,watzinger2018germanium,hendrickx2020single,hendrickx2021four} because of the strong intrinsic spin-orbit coupling, which is not present for electron spins. The intrinsic spin-orbit coupling allows rapid electrical manipulation of hole-spin qubits, without the need for additional bulky device features such as micro-magnets or ESR strip-lines. Further, the g-factor\cite{crippa2018electrical,venitucci2018electrical,liles2021electrical,froning2021strong} and spin-orbit coupling\cite{froning2021ultrafast,bosco2021hole} for holes are both tunable, providing a wide range of in-situ control over hole-qubits. In addition, hole-spins have the potential for enhanced coherence times due to suppressed hyperfine coupling\cite{prechtel2016decoupling}, and the potential for configuring decoherence sweet-spots by tuning the spin-orbit interaction\cite{wang2021optimal,bosco_hole_2021,mutter2021all,piot_single_2022}. 

\begin{figure*}[t]
	\centering
	\includegraphics[width=16cm]{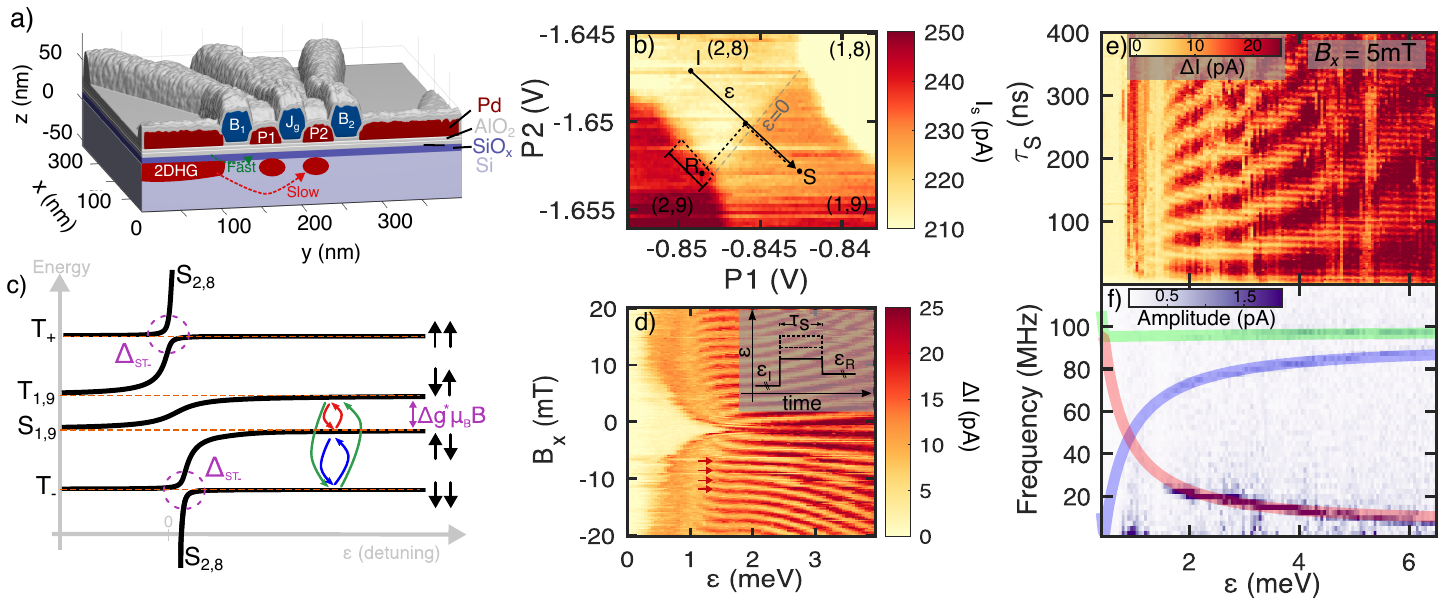}
	\caption[]{\small \textbf{Device operating point and energy spectrum}. 
	a) A 3D model of the device, showing a cross section through the double quantum dot region. A full SEM image is shown in Extended Figure \ref{fig:extfigure-1}a. The tunneling to the P1 dot was fast (<100~ns), while loading onto P2 dot is slow (>40~$\mu$s). This asymmetry in tunneling allows latched readout. b) Stability diagram of the (2,8)-(1,9) transition. The labels (N,M) indicate the number of holes in the P1 and P2 dot respectively. The colour scale is sensor current (I$_s$) in pA, and a full stability diagram down to (0,0) is presented in Extended Data Figure \ref{fig:extfigure-1}b. Zero detuning ($\epsilon=0$) is defined as the (2,8) and (1,9) charge degeneracy point, and positive detuning when the spins are separated into (1,9). Full details of the key points are discussed in the methods. c) Eigenstates calculated using the singlet-triplet Hamiltonian $H_{ST}$ defined in the methods. The coloured arrows indicate the energy transitions observed in the preceding experiments, and $\Delta_{ST}$ indicates the size of the avoided crossing between $\ket{S}$ and $\ket{T_\pm}$. Orange dashed lines show how the (1,9) states evolve in absence of spin-orbit coupling. d) The results of a spin funnel experiment with the pulse sequence shown in inset. The spin-funnel experiment was performed by initialising $\ket{S_{2,8}}$, followed by a rapid pulse to a point along the detuning axis ($\epsilon$). At each $B_x$ and $\epsilon$ the state was allowed to evolve for a fixed separation time, $\tau_{s}$=100~ns, followed by a pulse to the readout point. The change in sensor signal ($\Delta I$) due to this pulse indicates the likelihood of the returned state being singlet (low $\Delta I$) or triplet (high $\Delta I$) (described in more detail in the methods). Red arrows indicate the $\Delta g$-driven oscillations. e) The same pulse procedure as in d) except the magnetic field is fixed at B$_x$=5 mT and we investigate the effect of varying the separation time ($\tau_S$) at each detuning ($\epsilon$). f) The corresponding FFT at each detuning. Transparent lines indicated the best fit of the observed energy splittings to the eigenstates of $H_{ST}$, and the colours correspond to the transitions indicated in c).}
    \label{fig:Figure1}
\end{figure*} 

Despite the opportunities holes offer there currently are only limited studies of hole based singlet-triplet qubits. Recently, a hole-spin singlet-triplet qubit was demonstrated in Ge\cite{jirovec2021singlet}, where the strong spin-orbit coupling resulted in non-trivial qubit dynamics\cite{jirovec2022dynamics}. However, the spin-orbit effects in Si devices vary from Ge devices\cite{sarkar2023electrical,wang2023electrical}, therefore similar investigations in silicon would provide valuable understanding of silicon hole-spin effects. Recent experiments in silicon FinFET's have revealed an anisotropic exchange coupling for holes due to the spin-orbit interaction\cite{geyer2022two}, which may provide unique functionalities for hole-spin singlet-triplet qubits. Indeed, theoretical predictions have suggested that the non-trivial relationship between spin-orbit coupling and the site dependent g-tensors may allow hole-spin singlet-triplet qubits to avoid leakage errors\cite{mutter2021all}. However to-date there are no demonstrations of a singlet-triplet qubit using holes in silicon.

In this work we demonstrate a hole-spin singlet-triplet qubit formed in a planar MOS silicon double quantum dot. The planar structure provides a platform suitable for scaling up to the large arrays of coupled qubits needed for quantum circuits and error correction\cite{fowler2012surface,hill2015surface,li2018crossbar}. Additionally, the planar layout enables the straightforward implementation of a charge sensor. Using this charge sensor, we identify the exact hole occupation of the double dot, which is critical for experimental reproducibility and detailed theoretical modelling of this system. In addition to characterising the key parameters of the qubit, we perform an investigation into the anisotropy of the two-hole eigenstates. By comparing the experimental results with a model that includes spin-orbit coupling and anisotropic site dependent g-tensors, we identify key features in the eigenstates that allow the improvement of the initialisation fidelity and reduction in the readout errors.

\subsection*{Device and operating regime}
The hole-spin singlet-triplet qubit is formed using a planar-silicon double quantum dot device, fabricated using industrially compatible CMOS techniques. Figure \ref{fig:Figure1}a) shows a model 3D cross section of the double quantum dot region. Multilayer palladium gates define the double quantum dot with P1 and P2 operating as plunger gates, while J$_g$ provides in-situ control of the interdot tunnel coupling $t_c$\cite{jin2023combining}. The device employs ambipolar charge sensing\cite{de2020ambipolar}, with an adjacent $n$MOS SET allowing the absolute charge occupation of each quantum dot to be determined.

Figure \ref{fig:Figure1}b) shows a stability diagram measured using the charge sensor. We perform all measurements in the (2,8)-(1,9) configuration, which is equivalent to a (2,0)-(1,1) spin system due to orbital shell filling\cite{liles2018spin}. We initialised singlet states by dwelling deep in (2,8) where $\ket{S_{(2,8)}}$ is the lowest energy eigenstate (point I). Manipulation of the state was performed by pulsing to a position along the detuning axis ($\epsilon$) and dwelling there for a variable time $\tau_s$. Readout of the state was performed by pulsing to point R (following the dashed trajectory), where latched Pauli-Spin-Blockade\cite{studenikin2012enhanced,harvey2018high} readout allowed identification of either the blocked triplet or the unblocked singlet states based on the average sensor current (see methods and Extended Data Figure \ref{fig:extfigure-1}).

\begin{figure*}[t]
	\centering
	\includegraphics[width=16cm]{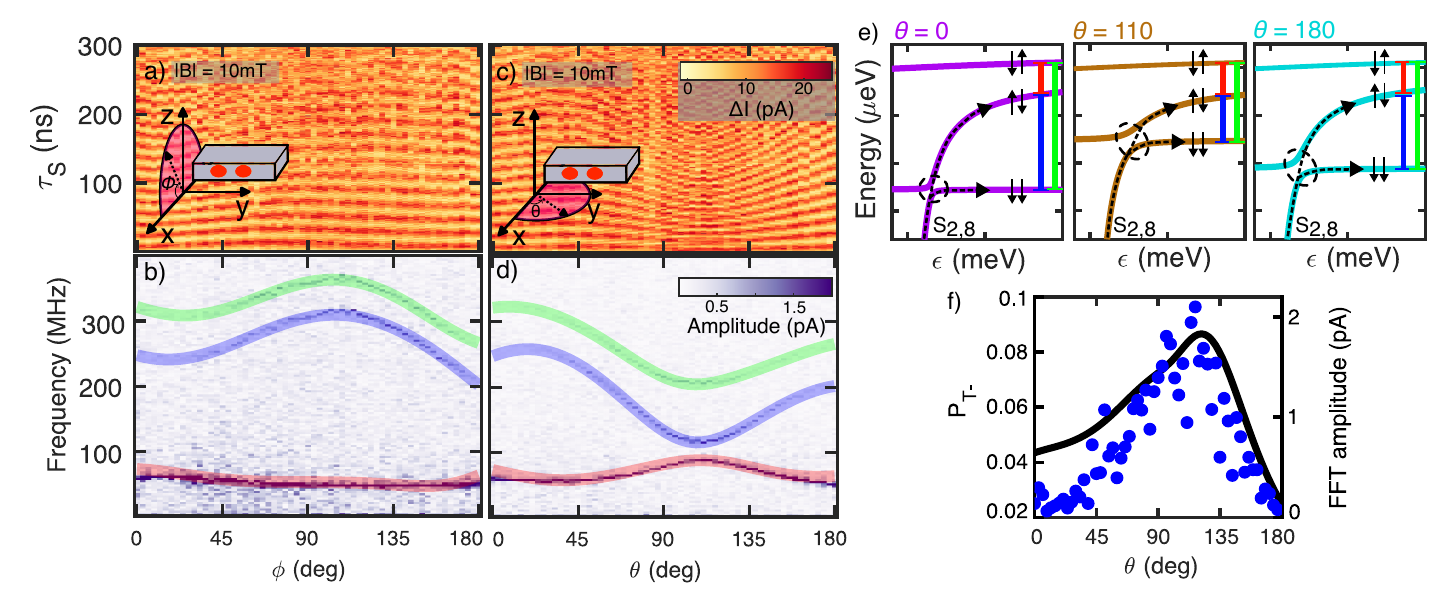}
	\caption[]{\small \textbf{Eigenenergy anisotropy with respect to magnetic field orientation}. 
	 a-b) Shows the sensor signal and resulting FFT when using the pulse sequence given in Fig \ref{fig:Figure1}d) (Initialise-Separate-Readout). Detuning is fixed at $\epsilon$ = 1.9 meV and a 10 mT magnetic field is rotated by 180$^\circ$ through the x-z plane. c-d) Shows the same experiment for a rotation through the sample x-y plane. Transparent solid lines show the optimal fit of $H_{ST}$ to the experimental data. A zoom in of the first 60~ns of c) is shown in Extended Data \ref{fig:extfigure-Model}a), highlighting the multiple frequencies present. See Supplement section S2.B for full 360$^\circ$ data set in the x-y plane. e) Shows the eigenenergies for $|B|$~=~10~mT magnetic field is applied at $\theta$~=~0$^\circ$ (purple), 110$^\circ$ (brown), and 180$^\circ$ (cyan) in the x-y plane respectively. The y-axis ticks are in 0.5~$\mu$eV, and x-axis ticks are separated by 1~meV. The energy splitting corresponding to the three FFT peaks in d) are indicated by the red, blue and green vertical lines. The size and location of the $\ket{S}$-$\ket{T_-}$ avoided crossing varies with field orientation (black dashed circle), resulting in anisotropy in the Landau-Zener transition probability between $\ket{S_{2,8}}$$\rightarrow$$\ket{T_-}$ during the ramp-in/ramp-out. The black dashed lines indicated the splitting of initial state when pulsing across the $\Delta_{ST}$ avoided crossing. f) The solid black line is the calculated probability of the $\ket{S_{2,8}}$ loading into $\ket{T_-}$ during the separation pulse ($P_{T_-}$, left axis). Blue markers indicate the amplitude of the $\ket{S}$$\leftrightarrow$$\ket{T_-}$ FFT peak in d) (transparent blue). Both data are plotted as a function of in-plane magnetic field angle. The trend in $P_{T_-}$ correlates with the amplitude of the $\ket{S}$$\leftrightarrow$$\ket{T_-}$ oscillations in d). The is correlation between $P_{T_-}$ and the $\ket{S}$$\leftrightarrow$$\ket{T_-}$ oscillation amplitude is expected since the $\ket{S}$$\leftrightarrow$$\ket{T_-}$ oscillations are enhanced as the probability of loading the $\ket{T_-}$ state increases.
  }
	\label{fig:Figure2}
\end{figure*} 

\subsection*{System Hamiltonian and eigenenergies}
To model the two-spin system we consider a 5$\times$5 Hamiltonian, $H_{ST}$, which includes Zeeman, spin-orbit and orbital terms. The full details of the two-hole singlet-triplet Hamiltonian are provided in Supplementary section S1. For the Zeeman Hamiltonian, we include independent 3x3 symmetric g-tensors for the left and right dot, $\tensor{g}_L$ and $\tensor{g}_R$ respectively. Hole-spins in silicon are known to have strongly anisotropic g-tensors\cite{crippa2018electrical,liles2021electrical,geyer2022two}, where variations in the g-tensor are produced by non-uniform strain\cite{liles2021electrical,abadillo2022hole}, spin-orbit coupling\cite{sen2023classification} and differences in the confinement profile between the two dots\cite{ares2013nature}. Hence, we do not assume that $\tensor{g}_L$ and $\tensor{g}_R$ are correlated or share the same principle spin-axes. For the spin-orbit Hamiltonian we include a spin-orbit vector, $\Vec{t}_{so} = (t_x,t_y,t_z)$, parameterising the effect of spin-orbit coupling in the laboratory reference frame indicated in Figure \ref{fig:Figure1}a).

In Figure \ref{fig:Figure1}c) we plot the eigenenergies of $H_{ST}$ as a function of detuning. At negative detuning the eigenstates are the ($\ket{T_+},\ket{T_0},\ket{T_-},\ket{S},\ket{S_{(2,8)}}$) basis states. At large positive detuning the eigenstates evolve into the $\ket{S_{2,8}}$ state and the four two-spin states ($\ket{\uparrow \uparrow},\ket{\uparrow \downarrow}, \ket{\downarrow \uparrow}, \ket{\downarrow \downarrow}$), which are defined by the sum or difference of the Zeeman energy in the two dots. The $\ket{\uparrow \downarrow}$ and $\ket{\downarrow \uparrow}$ eigenstates have energy splitting given by 
\begin{align}
    E_{ST_0} &= \sqrt{J(\epsilon , t_c)^2 + \Delta E_z^2}
    \label{eqn:Est}
\end{align}
where 
\begin{align*}
    J(\epsilon)&= \sqrt{\frac{\epsilon^2}{4}+2t_c^2}-\frac{\epsilon}{2}\\
    \Delta E_z &=|\Delta g^*| \mu_B |\vec{B}|
\end{align*}
$\epsilon$ is the detuning energy, $t_c$ is the interdot tunnel coupling, $\Delta g^*$ is the difference in the effective g-factors for the applied magnetic field vector $\vec{B}$ (see Supplement S1), and $\mu_B$ is the Bohr magneton. Since strong spin-orbit coupling results in an anisotropy in $|\Delta g^*|$ with respect to magnetic field orientation, we expect $E_{ST_0}$ to exhibit a non-trivial anisotropy\cite{geyer2022two}. An avoided crossing occurs between the $\ket{S_{2,8}}$ and $\ket{T_\pm}$, with the amplitude of the avoided crossing ($\Delta_{ST_\pm}$) determined by the interplay between the spin-orbit vector and the difference in the projection of the g-tensors for the given field orientation\cite{jirovec2022dynamics}. 

Figure \ref{fig:Figure1}d) shows the charge sensor response of a spin-funnel experiment used to characterise the singlet-triplet system\cite{petta2005coherent}. The spin-funnel experiment was performed by allowing a singlet state to time-evolve for $\tau_{s}$~=~100~ns at each $B_x$ and $\epsilon$. The change in sensor signal ($\Delta I$) then measures the likelihood that the final state is either singlet (low $\Delta I$) or triplet (high $\Delta I$). A clear funnel edge is visible in $\Delta I$ when the detuning point coincides with the $\ket{S}$ and $\ket{T_\pm}$ avoided crossing\cite{petta2005coherent}. In addition, on the positive detuning side of the funnel edge we see oscillations that result from $\Delta g$-driven $\ket{S}$$\leftrightarrow$$\ket{T_0}$ oscillations (red arrows). 

Figure \ref{fig:Figure1}e) we demonstrate the time evolution of the singlet at each detuning. The experimental procedure is the same as Figure \ref{fig:Figure1}d), however here we varied the separation time ($\tau_S$) at each detuning ($\epsilon$) and held the magnetic field constant at $B_x$~=~5~mT. Figure \ref{fig:Figure1}f) shows the FFT of $\Delta I$ at each $\epsilon$, revealing three clear oscillation frequencies, each with a distinct detuning dependence. Each oscillation frequency results from mixing between the three lowest eigenstates at the separation detuning ($\epsilon$). The lowest frequency (red) results from oscillations between $\ket{S}$$\leftrightarrow$$\ket{T_0}$ states, the middle frequency (blue) results from oscillations between $\ket{S}$$\leftrightarrow$$\ket{T_\pm}$, and the highest frequency (green) results from oscillations between $\ket{T_0}$$\leftrightarrow$$\ket{T_\pm}$. The corresponding transitions are indicated by coloured arrows in Figure \ref{fig:Figure1}c) and a full description is provided in the methods.

We fit the observed frequencies in Figure \ref{fig:Figure1}f) to the eigenergies of the singlet-triplet Hamiltonian, $H_{ST}$ and extract key parameters of the two-hole system. Transparent lines in Figure \ref{fig:Figure1}f) show the best fit, demonstrating good agreement between the observed and theoretical eigenenergies. Based on the best fit we extract $t_c = 9\pm1$~$\mu$eV and two effective g-factors of $0.8\pm0.1$ and $1.2\pm0.1$ for $B_x$.

\subsection*{Anisotropic g-tensors and spin-orbit coupling}

To characterise the key parameters of the two-hole system we investigate the effect of magnetic field orientation on the two-hole eigenenergies. Figure \ref{fig:Figure2}a) shows $\Delta I$ as a function of $\tau_{S}$ for a range of magnetic field orientations in the x-z plane and Figure \ref{fig:Figure2}b) shows the resulting FFT of $\Delta I$. Figures \ref{fig:Figure2}c-d) repeat the same experiment for a rotation of the magnetic field through the x-y plane. Clear anisotropy with respect to magnetic field orientation can be observed, which results from the interplay between spin-orbit coupling and the orientation of the g-tensors. The visibility of the higher frequency (blue and green) oscillations also shows a strong dependence on the magnetic field orientation. In particular, the FFT amplitude of the higher frequency (blue and green) oscillations is suppressed for $B_x$ and enhanced for $B_y$ and $B_z$.  

The 3x3 g-tensors for each dot and the spin-orbit orientation can be extracted by fitting the data in Figure \ref{fig:Figure2}a-d) to the eigenenergies of $H_{ST}$. The fitting procedure is discussed in Supplementary sections S3-S5. The transparent lines in Figures \ref{fig:Figure2}b) and d) indicate the frequency of the respective FFT peaks for the optimal fit parameters. For the optimal fit we find ($t^x_{so}, t^y_{so}, t^z_{so}$)~=~($-37~\pm~2,107~\pm~4,0~\pm~20$)~neV, giving $|\vec{t}_{so}|$ = 0.12~$\mu$eV. Notably, $\vec{t}_{so}$ is oriented in-plane with the 2DHG, consistent with expectations for heavy holes in planar silicon\cite{winkler2003spin}. Further, the in plane spin-orbit vector has components in both $t^x_{so}$ and $t^y_{so}$, indicating that a combination of Rashba (oriented perpendicular to the double dot axis) and Dresselhaus (oriented parallel to the double dot axis) spin-orbit components are present\cite{hung2017spin}. The full g-tensors are presented in the Extended Data Figure \ref{fig:extfigure-5}. We find that the orientation of the g-tensor principle axes for left and right dots are slightly misaligned, which may result from differences in confinement profile or non-uniform strain between the left and right dot. The observation of misalignment in the g-tensor principle axes suggests accurate modelling of multiple quantum dot systems in silicon should incorporate site dependent g-tensors with differing principle axes.

The anisotropy in the FFT amplitudes in Figure \ref{fig:Figure2}a-d) is caused by the probability of transitioning from $\ket{S_{2,8}}$ into $\ket{T_-}$ during the pulse from (2,8) to (1,9).  When pulsing from (2,8) to (1,9) the $\Delta_{ST_\pm}$ avoided crossing causes the initial $\ket{S_{2,8}}$ state to be split between $\ket{S}$ and $\ket{T_-}$ with a ratio determined by the Landau-Zener transition probability\cite{petta2010coherent} (see methods). Larger $\Delta_{ST_\pm}$ favours $\ket{T_-}$ states, while smaller $\Delta_{ST_\pm}$ favours $\ket{S}$. In Figure \ref{fig:Figure2}e) we plot the energy spectrum of the two-hole system for various in-plane magnetic field orientations. The magnetic field orientation strongly influences the magnitude of $\Delta_{ST_\pm}$ and the position in detuning ($\epsilon_{\Delta}$) at which the avoided-crossing occurs. As a result, the magnetic field orientation impacts the likelyhood of populating the $\ket{T_-}$ state during the separation ramp and thus impacts the amplitude of the $\ket{S}$$\leftrightarrow$$\ket{T_-}$ FFT peak.

We simulated the experimental pulse sequence using QuTiP\cite{johansson2012qutip} and calculated the $\ket{S_{2,8}}$$\rightarrow$$\ket{T_-}$ transition probability ($P_{T_-}$), which yielded good agreement with the measured FFT amplitude. In Figure \ref{fig:Figure2}f), the solid line shows the calculated $P_{T_-}$ using the optimal fit parameters for a range of in-plane magnetic field orientations (See Supplement section S3 for details). The circles in Figure \ref{fig:Figure2}f) show the observed amplitudes of the $\ket{S}$$\leftrightarrow$$\ket{T_-}$ FFT peaks from Figure \ref{fig:Figure2}d) (blue). The trend in $P_{T_-}$ matches the anisotropy in the measured amplitudes of the $\ket{S}$$\leftrightarrow$$\ket{T_-}$ FFT peaks from Figure \ref{fig:Figure2}d), with both exhibiting a peak around $\theta$~=~120$^\circ$, and an asymmetric reduction towards 0$^\circ$ and 180$^\circ$. The correlation between FFT amplitude and the calculated $P_{T_-}$ demonstrates that the model $H_{ST}$ and optimal fit parameters captures the dynamics of the hole-spin qubit well (See Extended Data Figure \ref{fig:extfigure-Model}a).

We now consider how hole-spin singlet-triplet qubits can be further optimised by using the anisotropic response of the system to a magnetic field. For the singlet-triplet qubit studied here, optimal initialisation protocol would suppress the likelyhood of the $\ket{S_{2,8}}$ state loading into the $\ket{T_{-}}$ leakage state. In the presence of large spin-orbit coupling and/or large $\Delta g$, even rapid separation pulses may be unable to satisfy the non-adiabatic Landau-Zener requirement imposed by the large $\Delta_{ST_-}$ avoided crossing. However, with knowledge of the orientation of spin-orbit vector ($\vec{t}_{so}$), and the g-tensors, we can identify optimum field orientations to minimise $\Delta_{ST_-}$ and thus enhance initialisation fidelity. Indeed, in Figure \ref{fig:Figure2}f) we have shown that the B$_{x}$ magnetic field suppress loading into the $\ket{T_{-}}$ state, and therefore is an optimal field orientation for the singlet-triplet qubit in this system.

\subsection*{Coherent $\Delta$g-driven oscillations}

\begin{figure*}[t]
	\centering
	\includegraphics[width=15cm]{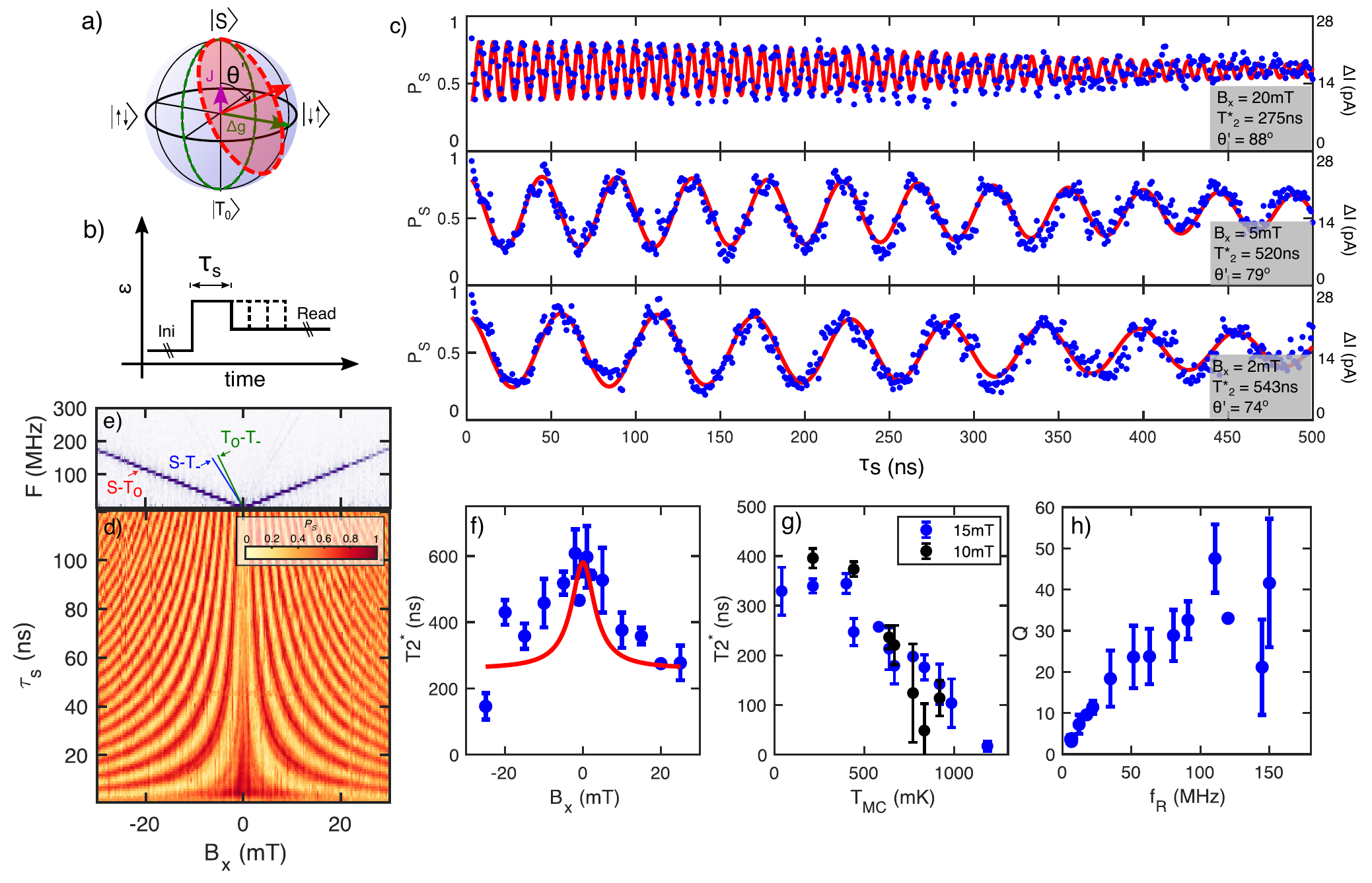}
	\caption[]{\small \textbf{$\Delta g$-driven coherent oscillations}. 
	a) Shows a schematic of the Bloch sphere for the singlet-triplet qubit. The Zeeman energy difference $\Delta E_z$ produces oscillations about the x-axis ($\ket{S}$$\leftrightarrow$$\ket{T_0}$ oscillations), while exchange coupling $J$ produces oscillations about the z-axis (ie $\ket{\uparrow\downarrow}$$\leftrightarrow$$\ket{\downarrow\uparrow}$ oscillations). When both $J$ and $\Delta E_z$ are non-zero, the angle of rotation with respect to the Bloch sphere z-axis is given by $\theta'$ = arctan$(\frac{\Delta E_z}{J})$ as indicated by the red trajectory. b) Shows the pulse sequence used to drive the $\Delta g$-driven oscillations. We initialised in $\ket{S_{2,8}}$, then applied a rapid separation pulse along the detuning axis. The system was then held at a fixed positive detuning ($\epsilon$) for a separation time ($\tau_S$), before performing readout. c) Shows the observed $\ket{S}$$\leftrightarrow$$\ket{T_0}$ oscillations as a function of separation time, for $B_x$ = 20~mT, 5~mT and 2~mT. The solid red line is the best fit of Eqn. \ref{eqn:Osc} to the data. The singlet probability, $P_S$, at each $\tau_S$ is extracted from the change in the normalised sensor signal $\Delta I$ (shown on right axis, see methods for full details). d) Shows the $\ket{S}$$\leftrightarrow$$\ket{T_0}$ oscillations over a field range of $B_x$~=~$\pm$30~mT. e) Shows the FFT of the oscillations observed in d). A linear increase in the oscillation frequency is expected since $hf$~=~$\Delta E_{ST_0}$ (Eqn. \ref{eqn:Est}). The $B_x$ magnetic field orientation was used for these experiments since it results in the least leakage into $\ket{T_-}$ (Figure \ref{fig:Figure2}). However, residual loading of the $\ket{T_-}$ leakage state results in weak $\ket{S}$$\leftrightarrow$$\ket{T_-}$ (blue) and $\ket{T_0}$$\leftrightarrow$$\ket{T_-}$ (green) oscillations for $B_x$. f) Shows the $T2^*$ for different magnetic fields, where the solid line is the best fit of Eqn. \ref{eqn:T2Fit} to both this data and the data in Figure \ref{fig:Figure4}e). g) Shows the effect of mixing chamber temperature on $T_2^*$ for two different magnetic fields. h) The qubit quality factor ($Q$) as a function of the $\Delta$g-driven oscillation frequency ($f_R$), measured at $T_{MC}$=30~mK. All data in Figure \ref{fig:Figure3} was collected with $J_g$~=~1.2~V.
    }
	\label{fig:Figure3}
\end{figure*} 

We now turn to experiments to characterise the hole-spin qubit. Here, the qubit is defined using the $\ket{S}$ and $\ket{T_0}$ states of the double quantum dot. The simplified Hamiltonian for this system can be written as
\begin{equation}
    H_{ST} = \frac{J}{2} \sigma_z + \frac{\Delta E_z}{2} \sigma_x
\end{equation}
where $J$ defines the exchange energy, $\Delta E_z = |\Delta g^*| \mu_B |B|$, $|B|$ is the magnitude of the applied field, $\sigma_{x,z}$ are the respective Pauli matricies and $\mu_B$ is the Bohr magneton. The Bloch sphere for this qubit system is shown in Figure \ref{fig:Figure3}a). Rotations around the Bloch sphere can be driven by controlling $J$ and $\Delta E_z$ at the separation point\cite{hanson2007universal,jock2018silicon}, and Figure \ref{fig:Figure3}b) shows a schematic of the pulse sequence used.

Figure \ref{fig:Figure3}c) plots the measured singlet probability $P_S$ as a function of separation time $\tau_S$ for three different |$B_x$|, demonstrating oscillations in $P_S$. Solid lines in Figure \ref{fig:Figure3}b) show the best fit of the data to the equation 
\begin{equation}
    P_S = A \text{cos}(2\pi f_R \tau_s+ \phi) \text{Exp}[ (\frac{\tau_s}{T^*_2})^\alpha] + C
    \label{eqn:Osc}
\end{equation}
where $f_R$ is the Rabi frequency, $\tau_s$ is the separation time, $\phi$ is a phase offset, $T_2^*$ is the qubit dephasing time, A is the oscillation amplitude, C is an offset, and $\alpha$ captures the noise colour (see Supplementary section S2.A).  

Analysis of the $\ket{S}$$\leftrightarrow$$\ket{T_0}$ oscillations over a range of $B_x$ was used to characterise the qubit control frequency ($f_R$) and the coherence time ($T^*_2$). Figures \ref{fig:Figure3}d) and e) show a colour map of the $\ket{S}$$\leftrightarrow$$\ket{T_0}$ oscillations and a corresponding FFT at each $B_x$. We resolve $\ket{S}$$\leftrightarrow$$\ket{T_0}$ oscillations up to 150 MHz at 30 mT and have observed up to 400~MHz at 80~mT (Extended data Figure \ref{fig:extfigure-2}). To extract $\Delta g$ and $J$ we fit $f_R$ for each $B_x$ to Eqn. \ref{eqn:Est}. The fit yields $J$~=~6~MHz at the separation point ($\epsilon = 1.9$ meV), and $|\Delta g|$~=~0.41 which is in agreement with the effective g-factors extracted from Figure \ref{fig:Figure1}f. In Extended data Figure \ref{fig:extfigure-2}d) we show electrical control over $\Delta g^*$ using the $J_g$ gate, with a trend of $d\Delta g^*/dJ_g\approx$~0.9~V$^{-1}$, demonstrating in-situ electrical control of the qubit control frequency.

We now show the decoherence in this qubit can be explained by fluctuations in both $\epsilon$ and $\Delta E_Z$ and we quantify their magnitudes. Figure \ref{fig:Figure3}f) shows $T_2^*$ for a range of $B_x$, where each $T_2^*$ has been extracted using a fit to Eqn \ref{eqn:Osc}. Dephasing is caused by fluctuations in the energy splitting between the $\ket{S}$ and $\ket{T_0}$ states\cite{dial2013charge} (Eqn \ref{eqn:Est}). Hence, the variation in $T_2^*$ can be modelled using
\begin{equation}
    \frac{1}{T_2^*} = \frac{\pi\sqrt{2}}{h}\sqrt{  (\frac{J}{E_{ST_0}} \frac{dJ}{d\epsilon} \delta \epsilon)^2 + (\frac{\Delta E_Z}{E_{ST}}\delta \Delta E_Z)^2 } 
    \label{eqn:T2Fit}
\end{equation}
where $\delta \epsilon$ is the noise in detuning and $\delta \Delta E_Z$ is the effective magnetic noise. A common fit of the data in Figures \ref{fig:Figure3}f) and  \ref{fig:Figure4}d) (discussed later) was used to extract $ \delta \epsilon~=~30~\mu$eV and $ \Delta E_Z = 4 $~neV. These values are comparable to those previously reported for electrons in silicon micro-magnet devices\cite{wu2014two} and for holes in planar Ge devices\cite{jirovec2021singlet}. The similarity in $\delta \epsilon$ between holes in planar Si and planar Ge suggests that the highly disordered SiO$_2$ oxide does not significantly enhance the effect of charge noise compared to planar Ge heterostructures, where the quantum dot is buried tens of nanometers below the surface. Further, the similarity in $\delta \epsilon$ between this work in $p$MOS silicon and studies of electrons in $n$MOS silicon\cite{wu2014two} suggests that the level of charge noise is not impacted by the polarity of the gate bias. 

In Figure \ref{fig:Figure3}g) we plot $T_2^*$ of the $\Delta$g-driven oscillations as a function of the fridge mixing chamber temperature ($T_{MC}$). $T_2^*$ is approximately independent of $T_{MC}$ up to 400 mK ($k_BT=34$ $\mu$eV ), where $T_2^*$ begins to drop. The $T_2^*$ behaviour shows the same trend at $B_x$ = 15 mT ($\Delta E_{ST_0} = 0.35$ $~\mu$eV) and $B_x$ = 10~mT ($\Delta E_{ST_0} = 0.23$~$\mu$eV). Interestingly, we find that the noise colour ($\alpha$) appears to whiten as the temperature of the fridge increases, consistent with recent experiments in other hole spin qubits\cite{camenzind2022hole} (see Supplement section S2.A).

Finally, in Figure \ref{fig:Figure3}h) we show that the quality factor ($Q = f_R \times T_2^*$) of the $\Delta$g-driven singlet-triplet oscillations increases as the control speed increases. The quality factor quantifies the number of coherent oscillations that can be completed within the coherence time. A promising feature of this qubit is that Q increases with increasing $f_R$, indicating that the qubit control speed can be increased without degrading the quality. 

\subsection*{Coherent exchange-driven oscillations}
\begin{figure*}[hbt]
	\centering
	\includegraphics[width=15cm]{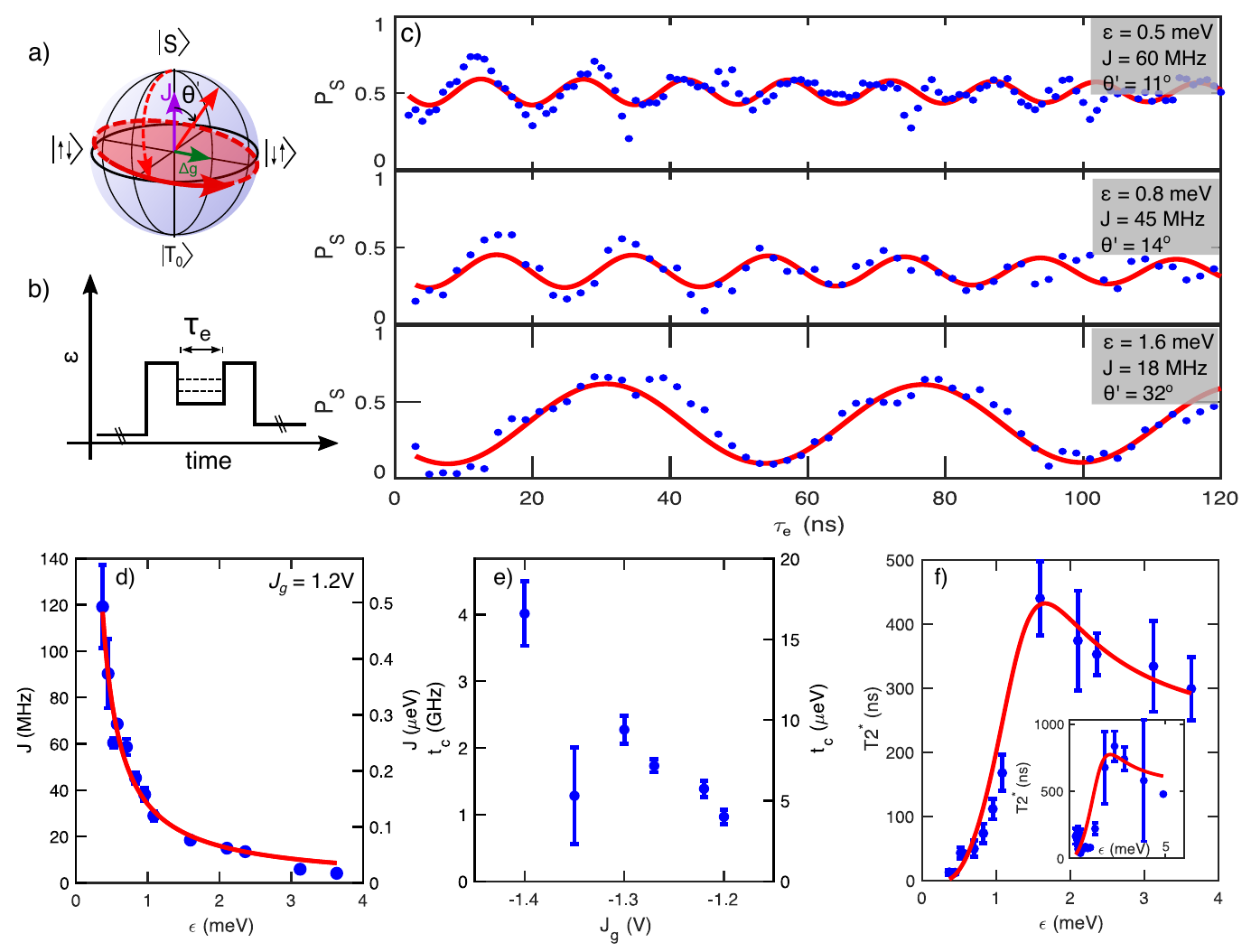}
	\caption[]{\small \textbf{Exchange-driven coherent oscillations}. 
	a) A schematic of the state evolution and b) shows the pulse sequence used to achieve exchange-driven oscillations. A $\Delta g$-driven $\pi/2$ pulse brings the qubit to the equator (red dashed trajectory), then a rapid pulse to low detuning is used to suddenly increase J, changing the angle of rotation, resulting in exchange-driven oscillations around the equator (red trajectory). The orientation of the oscillation is tilted by $\theta'$ from the z-axis. c) Exchange-driven oscillations for three different detunings ($\epsilon$= 0.5~meV, 0.8~meV, and 1.6~meV). Experiments were performed with a fixed magnetic field of B$_x$~=~2~mT, such that a $\Delta g$-driven $\pi/2$ pulse takes 65ns. The solid line shows a best fit to Eqn. \ref{eqn:Osc}, allowing extraction of J and $T_2^*$ at each $\epsilon$. d) Exchange energy as a function of detuning, measured using the exchange pulse. The solid red lines shows the fit to $J(\epsilon , t_c)= \sqrt{\frac{\epsilon^2}{4}+2t_c^2}-\frac{\epsilon}{2} $, allowing the extraction of the tunnel coupling $t_c$. e) Tunnel coupling extracted for a range of different $J_g$ gate voltages, where each $t_c$ is extracted from the $\epsilon$ dependence of the exchange-oscillation frequency. f) Dephasing time $T_2^*$ as a function of the detuning. The solid red line is a joint fit of Eqn. \ref{eqn:T2Fit} to this data and the data in Figure \ref{fig:Figure3}f). All data in c), d) and f) was collected for $J_g$~=~1.2~V, and the inset in f) shows the dephasing time as a function of the detuning for J$_g$~=~-1.4~V.
	}
	\label{fig:Figure4}
\end{figure*} 
In order to achieve full control of the singlet-triplet qubit it is necessary to produce control pulses around two orthogonal axes of the Bloch sphere. We use exchange-driven oscillations to rotate the qubit around the z-axis of the Bloch sphere. By combining $\Delta g$-driven (x-axis) and exchange-driven (z-axis) rotations, it is possible to achieve full control over the qubit Bloch sphere. 

Figures \ref{fig:Figure4}a) and \ref{fig:Figure4}b) show the experimental procedure for exchange-driven oscillations. A separation pulse with calibrated $\tau_S$ is first used to perform a $\Delta g$-driven $\frac{\pi}{2}$ rotation to bring the state to the equator of the Bloch sphere. A rapid pulse to low detuning is then applied, which suddenly increases J, producing a change in qubit rotation axis. The system is held at the exchange-point for $\tau_E$ to drive oscillations around the z-axis, then a second $\Delta g$-driven $\frac{\pi}{2}$ rotation is applied, followed by a pulse to the readout position. 

\begin{figure*}[t]
	\centering
	\includegraphics[width=6cm]{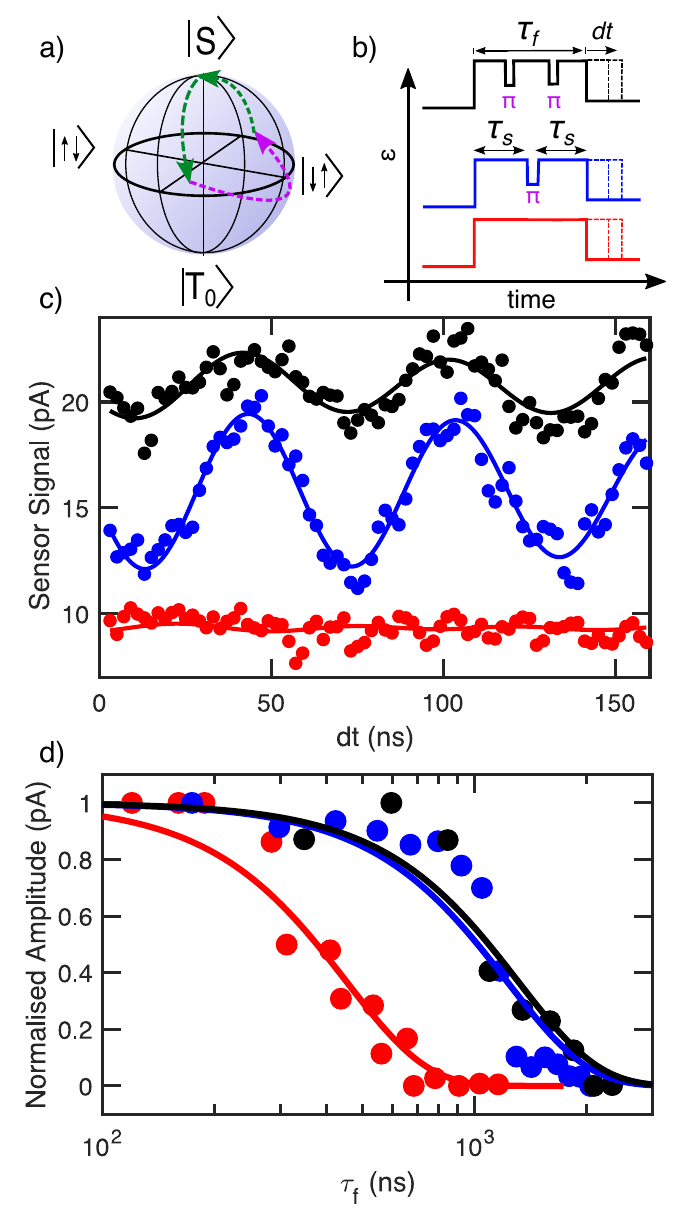}
	\caption[]{\small \textbf{Spin echo measurement at $B_x$ = 1.6 mT}. 
	a) Shows an example trajectory of the qubit-state for the refocusing pulse schematic of b). Free evolution driven by $\Delta g$ is allowed for a period of time $\tau_S(n) = (2n + 1/2)t_{\pi}$ (green trajectory), such that after any $\tau_S(n)$ the qubit will be at the equator. A refocusing pulse is incorporated as a $\pi$ exchange pulse, followed by a second period of free evolution for $\tau_S(n)$. The total time to perform this sequence is the free evolution time, $\tau_f$, which can be varied by increasing $n$, or repeating the cycle to include multiple refocusing pulses. The full pulse sequence results in the qubit refocusing to $\ket{S}$ after $\tau_f$. c) Residual $\ket{S}-\ket{T_0}$ oscillations after a free evolution time ($\tau_f$) of 1000~ns for no refocusing pulses (red), one refocusing pulse (blue) and two refocusing pulses (black). d) Normalised peak-to-peak amplitude of the residual $\ket{S}-\ket{T_0}$ oscillations as a function of free evolution time. For one (blue) and two (black) refocusing pulses the oscillations are clearly extended compared with the data for no refocusing pulse. The amplitude is normalised against the shortest free evolution for each data set in order to account for the fidelity of the $\pi$ exchange pulse. We extract $T^*_{Echo}$ based on a fit to the decay envelope of Eqn. \ref{eqn:T2Fit}, using $\alpha=2$ (see Supplement section S2.A). The experiments were performed for $B_x$ = 1.6 mT. 
	}
	\label{fig:Figure5}
\end{figure*} 

Figure \ref{fig:Figure4}c) demonstrates exchange-driven oscillations at three different detuning position ($\epsilon$). Reducing $\epsilon$ at the exchange position increases the exchange energy J, so that the angle of rotation tends towards 0$^\circ$ with respect to the Bloch sphere z-axis (since J~$\gg \Delta g \mu_B B$). In Figure \ref{fig:Figure4}d) we plot $J$ as a function of $\epsilon$ at the exchange point. The solid line shows the best fit of $J(\epsilon,t_c)$, allowing the extraction of the tunnel coupling (see Figure caption). This experiment was repeated for a range of $J_g$ gate voltages, and the resulting dependence of the tunnel coupling ($t_c$) on the $J_g$ gate voltage is shown in Figure \ref{fig:Figure4}e), demonstrating smooth control of $t_c$. Therefore, the exchange-driven oscillations are highly tunable, since $J$ can be electrically tuned either by varying $\epsilon$ with the plunger gates, or by tuning $t_c$ using the $J_g$-gate. These results demonstrate coherent exchange-driven z-axis control of the singlet-triplet qubit.

Figure \ref{fig:Figure4}f) presents $T^*_2$ as a function of $\epsilon$, which is used to characterise the coherence time of the exchange oscillations. The solid line shows the best fit to Eqn. \ref{eqn:T2Fit}, obtained by jointly fitting Figure \ref{fig:Figure4}e) and Figure \ref{fig:Figure3}f). The trend in $T^*_2$ is well explained by Eqn. \ref{eqn:T2Fit}, where charge noise ($\delta \epsilon$) dominates at low detuning due to enhanced $dJ/d\epsilon$, while Zeeman noise ($\delta \Delta E_Z$) dominates at large detuning where $dJ/d\epsilon \xrightarrow{}0$.

\subsection*{Spin echo measurement}

Finally, we investigate the use of spin-refocusing to enhance the qubit coherence time. Given that J and $\Delta E_Z$ are non-zero, rotations around the Bloch sphere occur at some angle offset from the pure x-axis or z-axis. Since the qubit trajectory is not solely around the x-axis or z-axis, the precise form of the refocusing pulse will vary as the qubit evolves. Therefore, complicated pulse engineering is required for perfect refocusing pulses\cite{wang2012composite}. Here, we implement a simplified procedure\cite{jirovec2021singlet}, which employs a $\pi$ rotation using an exchange pulse to enhance the observed coherence analogous to a Hahn echo. Figures \ref{fig:Figure5}a) and b) show the qubit evolution and pulse sequence for the refocusing procedure respectively. The spin echo experiment allowed the qubit to freely evolve for a time $t_f$, during which $N$ exchange-driven $\pi$ rotations are carefully interlaced to provide the refocusing echo. Full details of the refocusing procedure are provided in the methods.  

We demonstrate the enhancement of the qubit coherence by observing the residual $\Delta g$-driven singlet-triplet oscillations after the free evolution time, $t_f$. Figure \ref{fig:Figure5}c) shows the singlet-triplet oscillations after $t_f$ = 1000~ns for zero (red), one (blue) and two (black) refocusing pulses. When zero refocusing pulses are applied the singlet-triplet oscillations are completely lost after 1000~ns of free evolution. However, with refocusing pulses the singlet-triplet oscillations are visible even after $t_f > 1000$~ns. Figure \ref{fig:Figure5}d) shows the normalised amplitude of the residual $\Delta g$-driven singlet-triplet oscillations observed for a range of free evolution times. The application of one and two refocusing pulses clearly enhances the coherence of the qubit. We fit the decay of the peak amplitude to extract $T^{Echo}_2$ = 1220$\pm$150~ns for one refocusing pulse, and $T^{Echo}_2$ = 1300$\pm$200~ns for two refocusing pulses. When no refocusing pulses are applied we find $T^{Echo}_2$ = 550$\pm$50~ns, consistent with the measurements in Figure \ref{fig:Figure3}. Although we see an improvement of 120\% by applying one pulse, we see no significant improvement when using two refocusing pulses. This suggests the maximum $T^{Echo}_2$ may have been reached for this simplified refocusing procedure.

\subsection*{Conclusion}

In this work we have demonstrated a hole-spin singlet-triplet qubit in planar silicon. We demonstrate rapid $\ket{S}$$\leftrightarrow$$\ket{T_0}$ oscillations exceeding 400 MHz, two axis control via $\Delta g$-driven and exchange-driven oscillations, and enhancement of the qubit coherence time to >1~$\mu$s using spin-echo procedures. Developing a complete model of the energy spectrum provided insight into spin-qubit dynamics under rapid pulses across the $\ket{S}$$\leftrightarrow$$\ket{T_-}$ avoided crossing. The experimentally observed effects were well described by the model Hamiltonian, providing insight into methods to further optimise initialisation protocols in hole-spin qubits.

\small{

}

\section*{Methods}
\small
\subsection*{Sample details} 
The device was fabricated from high resistivity natural silicon. A SEM image of a nominally identical device is shown in Figure \ref{fig:extfigure-1}a) where the multi-layer Pd gate stack is achieved using 2nm ALD AL$_2$O$_2$ and a high quality 5.9nm SiO$_2$ gate oxide. This device combines nMOS and pMOS capabilities in order to implement a Single Electron Transistor (SET) as the charge sensor, while defining a hole double quantum dot\cite{de2020ambipolar,jin2023combining}. 

\subsection*{Device operation}
A Single Electron Transistor (SET) was defined by the ST (Sensor Top gate) and the left and right barrier gates (SLB and SRB)\cite{angus2007gate}. A double quantum dot was formed by carefully biasing the gates B1, P1, J$_g$, P2, and B2. Gates PR1 and PR2 were used to accumulate a 2D hole gas, which served as a reservoir of holes. In addition to providing confinement, the B1 and B2 gates can also modulate the tunneling rate between the hole quantum dots and the reservoirs. In this experiment we asymmetrically biased B1 and B2 (B1 = -0.6V, B2 = -0.4V). Configuring different lead-dot tunnel rates between the left and right dot enabled latched readout, which is discussed later. 

The SET acts as a highly sensitive charge sensor for the adjacent hole double quantum dot. In Figure \ref{fig:extfigure-1}b) we show a large stability diagram of the double quantum dot, measured by applying a small (1mV) pulse at 177Hz on P1 and monitoring the the charge sensor using a lock-in reference to this pulse\cite{liles2018spin}. In addition, dynamic feedback is used to maintain the SET at a sensitive position over the large range of plunger gates used\cite{yang2011dynamically}. The use of a charge sensor allows confirmation of the absolute hole occupation since the (0,0) region can be identified. We note that some transitions lose visibility (black dashed lines) since the tunnel rate becomes slower than the 177Hz used for the measurement. However all transitions can be inferred by the observation of avoided inter-dot transitions. The (2,8)-(1,9) region used in this experiment is indicated by a white circle. 

\subsection*{Experimental Setup}
All experiments were performed using a top loading BlueFors XLD dilution fridge with a 3-axis vector magnet. Unless otherwise stated the experiments were performed with the fridge at base temperature where the mixing chamber thermometer was 40 mK. Previous measurements indicated that at base this system achieves electron and hole temperatures of 120 mK. The device was fixed directly onto a brass sample enclosure that is thermally anchored to the probe cold-finger. GE varnish was used to mount the device directly onto a brass sample stage, and Al bond wires connect the sample to a home-made printed circuit board. All DC biasing was applied using a Delft IVVI digital-to-analogue-converter using lines which each pass through individual 50kHz low pass filters mounted to the cold finger (40 mK). The current through the SET was amplified using a Basel SP983c I-V preamplifier. DC currents were then monitored using a Keithly 2000, and standard low-frequency lock-in techniques were implemented using a SR830.

Rapid pulses were applied to gates P1 and P2 in order to perform initialisation, control and readout out of the singlet-triplet qubit. The pulses were applied using a Tabor WX1284 with 1.25 GS/s. Fast pulses from the WX1284 were routed to gates P1 and P2 using homemade RC bias-tees (R = 330~kOhm, C = 1.2~nF) on the sample PCB at the mixing chamber. The pulses were transmitted from room temperature to the circuit board using coaxial cables with 15~dBm of cold attenuation for thermalisation. 

\subsection*{Pulse sequence and readout}

Figure \ref{fig:extfigure-1}c) shows the stability diagram of the (2,8) - (1,9) region, which was used to for the singlet-triplet qubit. The data is Figure \ref{fig:extfigure-1}c) is obtained with no additional pulses on any gates, and shows the sensor DC current. The charge transitions and interdot transition are are indicated by dashed black lines. The rectangle indicated at the border of the (2,9)-(2,8) transition is a suitable region for performing spin-to-charge conversion via latched readout region (details below).

In Figure \ref{fig:extfigure-1}d) we show dashed lines reproducing the charge transitions from Figure \ref{fig:extfigure-1}c). In addition, key positions - `I', `S', `P' and `R' - are labelled. The solid lines between points `\textbf{I}nitialise', `\textbf{S}eparate', `\textbf{P}assing' and `\textbf{R}eadout' indicate the trajectory of pulses used to move between the points. The point `I' is used for initialisation of a S(2,8) state, which is achieved by dwelling at point `I' for 600~ns. The point `S' lies along the detnuing axis in (1,9) and is used to separate the two unpaired spins. For this experiment $t_c$ is sufficiently large that the pulses from I to S were approximately adiabatic across the (2,8)-(1,9) avoided crossing. The transitory point `P' is used to prevent unwanted pulsing across change transitions when returing to readout. The dwell time at `P' was 1~ns. As `P' is only a transitory point we omit it from pulse sequence descriptions (ie I-S-R implies I-S-P-R). Point `R' is chosen to allow latched PSB readout using the (2,8)-(2,9) charge transition (described below). Typically, the dwell time at `R' was 10-20~$\mu$s, such that the `R' comprises the majority of the pulse cycle. 

Figure \ref{fig:extfigure-1}d) shows the DC sensor current measured when sweeping the DC plunger gate voltages while applying the fixed I-S-R pulse sequence. A magnetic field of 2~mT is applied and a separation time of 51~ns was used to produce a high singlet-probability (calibrated to a 2$\pi$ rotation, see Figure \ref{fig:Figure2}c at 2~mT). The DC current is measured with an integration time much longer than the pulse cycle (integration time = 100 ms), such that the measured DC current is essentially the average current during the `R' stage. The DC current in Figure \ref{fig:extfigure-1}d) reproduces the charge transitions of Figure \ref{fig:extfigure-1}c) (where no pulses were applied), however rectangular region occurs at the border of the (2,9)-(2,8) charge transition. This feature is arises due to charge latched PSB as indicated in Figure \ref{fig:extfigure-1}e).
%

\subsection*{Singlet-triplet spin-to-charge conversion}
We take advantage of the latched readout to perform spin-to-charge conversion. The experiment uses a differential current method to determine the difference in DC current between a `measurement' pulse sequence and a `reference' pulse sequence. The `measurement' pulse sequence follows I-S-R for N repetitions. This sequence initialises into $\ket{S_{2,8}}$ at I, manipulates the qubit at S, then maps the resulting spin-state to either (1,9) or (2,8) charge-states at point 'R'. The `reference' pulse sequence follows a reverse pulse cycle, S-I-R. This cycle is used as a `reference' since regardless of the `S' parameter, the subsequent `I' stage initialises a singlet prior to readout. The number of repetitions, N, is chosen such that 1/2N~=~177~Hz, and a SR830 lock-in amplifier is then used to determine the difference in current between the measurement pulse cycle (I-S-R) and the reference pulse cycle (S-I-R). In this way the lock-in signal measures $\Delta I$ of the sensor, such that $\Delta I$~=~0 indicates the measurement pulse produced only singlets, while increasing $\Delta I$ > 0 indicates an increase in triplet probability. The maximum $\Delta I$ is related to the difference in I$_{DC}$ between the (2,8) and (2,9) regions (scaled by the pulse ratio). We have determined\cite{wu2014two} that 100\% triplet production would produce a maximum $\Delta I$ =~28 pA measured on the lock-in for Figures 3 and 4 of the main text, and this is used for the normalisation which allows conversion of average $\Delta I$ to singlet probability $P_S$. For all measurements the $\Delta I$ was obtained using a time constant of 0.3~ms, such that each $\Delta I$ is an average over approximately 15000 pulse cycles.

\subsection*{Origin of the three different oscillation frequencies}
Here we describe the origin of the three different oscillation frequencies observed in Figure \ref{fig:Figure1}f) and Figures \ref{fig:Figure2} a-d). The lowest frequency signal (red) arises from $\Delta g$-driven oscillations between $\ket{S}$$\leftrightarrow$$\ket{T_0}$ states. These $\Delta g$-driven oscillations occur since rapidly pulsing the $\ket{S_{2,8}}$ state to positive detuning produces a linear combination of the two-spin eigenstates ($\ket{\downarrow\uparrow}$ and $\ket{\uparrow\downarrow}$). Hence, after the rapid pulse to positive detuning, the time evolution results in singlet-triplet oscillations at a frequency defined by the energy difference $\Delta E_{ST_0}$ (Eqn. \ref{eqn:Est}).

The second frequency (blue) arises due to Landau-Zener oscillations between $\ket{S}$$\leftrightarrow$$\ket{T_\pm}$, which result from cycling the $\ket{S_{2,8}}$ between positive and negative detuning\cite{petta2010coherent}. The Landau-Zener oscillations occur since the separation pulse (rise time $\approx$4ns) is not fully diabatic across the $\ket{S}$$\leftrightarrow$$\ket{T_\pm}$ avoided crossing, resulting in finite probability of the $\ket{S_{2,8}}$ state following either the $\ket{\uparrow\downarrow}$ or the $\ket{\downarrow\downarrow}$ trajectories. During the separation time the a phase difference between the two trajectories accumulated. The readout pulse (fall time $\approx$4ns) traverses back across the avoided crossing, producing interference and resulting in a frequency defined by the energy splitting between the $\ket{\uparrow \downarrow}$ and $\ket{\downarrow \downarrow}$ eigenstates at the separation point. 

A weaker third frequency is observed (green) that occurs due to the combination of the two previous oscillations, at a frequency defined by the energy splitting $\Delta E = E_{\downarrow \uparrow} - E_{\downarrow \downarrow}$. As a result, the oscillation frequencies in Figure \ref{fig:Figure1}e) can be used to determine the eigenenergies of the system.

\subsection*{Landau-Zener transitions}
Here we discus the Landau-Zener process discussed in Figure \ref{fig:Figure2}. The probability of maintaining the initial eigenstate when ramping across an avoided-crossing is given by the Landau-Zener transition probability
\begin{equation}
    P_{LZ} = \text{exp}(\frac{-2\pi\Delta^2}{hv}),
\end{equation}
where $v = dE/dt$ is the energy level velocity, and $\Delta$ is the size of of the avoided crossing. For the two hole spin system we initialised a $\ket{S_{2,8}}$, then pulsed (ramp time $\approx$4~ns) to positive detuning. This pulse from negative to positive detuning traverses the $\Delta_{ST_\pm}$ avoided crossing. The magnetic field orientation strongly influences both the magnitude of $\Delta_{ST_\pm}$ and the position in detuning ($\epsilon_{\Delta}$) at which the avoided-crossing occurs. Therefore, the magnetic field orientation impacts the probability of transitioning from $\ket{S_{2,8}}$ into $\ket{T_-}$ during the pulse from (2,8) to (1,9). This in turn influences the amplitude of the $\ket{S}$$\leftrightarrow$$\ket{T_-}$ FFT peak (blue), since this peak amplitude is proportional to the probability of occupying  $\ket{T_-}$\cite{petta2010coherent}.

\subsection*{Extended characterisation measurements}
Figure \ref{fig:extfigure-2}a) shows the singlet-triplet oscillation frequency for a range of different in-plane magnetic fields ($B_x$). The solid line shows a fit of this data to the equation for $\Delta E_{ST}$ presented in the main text (Eqn. \ref{eqn:Est}).The best fit gives $|\Delta g^*| = 0.41\pm0.01$ and J = 6$\pm3$~MHz at the separation point ($\epsilon = 1.9$~meV). Figure \ref{fig:extfigure-2}b) shows the singlet-triplet oscillations measured out to a range of 100~mT in $B_x$. A FFT of the data is shown in Figure \ref{fig:extfigure-2}c) demonstrating singlet-triplet oscillations exceeding 400 MHz. 

We show electrical control over $\Delta g^*$, demonstrating that a key parameter of this hole-spin qubit can be electrically controlled in-situ. In Figure \ref{fig:extfigure-2}d) we plot the observed $\Delta g^*$ over a range of voltages applied to the $J_g$ gate. A trend of increasing $\Delta g$ as the $J_g$ gate becomes more positive was observed, with a slope of $d\Delta g^*/dJ_g \approx$ 0.9V$^{-1}$. Since the $J_g$-gate couples to each dot, it is not clear if both, or just one of the g-tensors are being electrically controlled. In addition to the electrical control of $\Delta g^*$, we also observed a temperature dependence of $\Delta g$. We observed a reproducible shift in $f_R$ of 6 MHz between mixing chamber temperatures of 30 mK and 1 K, corresponds to a shift in $\Delta g$ of 7\%. The origin of the temperature dependence is not clear, however temperature dependent shifts in the resonance frequency of spin qubits have recently been reported in other systems\cite{undseth2023hotter} (see Supplement section S3 for more details).

\subsection*{Spin echo measurement procedure}
Here we provide full details of the experimental procedure used for the spin echo measurement. This procedure was modelled on that developed for a planar Ge singlet-triplet hole-spin qubit\cite{jirovec2021singlet}. The $\ket{S_{2,8}}$ state was pulsed to large detuning, producing $\Delta g$-driven oscillations. We allow a dwell time of $t_{S}(n) = (2n + 1/2)t_{\pi}$, where $n$ is an integer and $t_{\pi}$ is the time for a $\Delta g$-driven $\pi$ rotation ($t_{\pi} \approx 60$ns). Hence, after any $t_S(n)$ the state should be at the equator of the Bloch sphere. After time $t_S(n)$, an exchange-driven $\pi$ rotation ($t_{\pi}^J \approx 7$ns) is applied. Following this we dwell at large separation for a second period $t_{S} = (2n + 1/2)t_{\pi}$. Multiple refocusing pulses can be added by repeating the above cycle. The blue schematic in Figure \ref{fig:Figure5}b) shows the sequence for one refocusing pulse, while the black schematic shows the sequence for two refocusing pulses. The free evolution time is the total time of the entire refocusing pulse sequence, $t_f = (2t_S(n) + t_{\pi}^J)N$, where N is the number of exchange driven $\pi$ pulses used. For a fixed number of exchange pulses (N), we can vary the free evolution time by increasing the $t_S(n)$ dwell time. The entire pulse sequence is designed to refocus the qubit to a singlet state, regardless of the total free evolution time $t_f(n)$. 

Given that the exchange pulses are shown to enhance coherence in Figure \ref{fig:Figure5}c), it is counter intuitive that the amplitude of the N=2 (black) oscillations are smaller than the N=1 (blue) oscillations in Figure \ref{fig:Figure5}c). However, this is due to the fidelity of the exchange pulse. The fidelity of the exchange pulse reduces the overall amplitude of the oscillations, which is the reason that the amplitude of the two-pulse data (black) is less than the one-pulse data (blue). The limited fidelity restricted the number of refocusing pulses we were able to apply, since for more than three refocusing pulses, the amplitude of the oscillations were reduced to the noise floor of the singlet-triple visibility, even for short $t_f$. As a result we plot the normalised oscillation amplitude in Figure \ref{fig:Figure5}c) to allow comparison between the zero, one and two exchange pulse measurements. 

\subsection*{Optimal fit for g-tensor and spin-orbit vector.}
We fit the frequencies of the three FFT peaks in Figure \ref{fig:Figure2}a) and Figure \ref{fig:Figure2}c) to predicted energy splitting in the singlet-triplet Hamiltonian $H_{ST}$ (defined in Supplement section S1). This allowed extraction of the g-tensors and spin-orbit vector for this hole-spin qubit. The spin-orbit vector (3 components) and the g-tensors (6 components each) result in a very large parameter spacem, with 16 parameters; $t_c$, $t_x$, $t_y$, $t_z$, 6 parameters for $g_L$ and 6 parameters for $g_R$. This presents a challenge since the large number of free parameters prevents identification of a unique fit to the FFT frequencies alone. In fact, we find more than 50 possible parameter combinations achieve giving good fits to the observed frequencies. 

We are able to effectively constrain the fits by simultaneously considering the eigenstate anisotropy and the state occupation probability indicated by the FFT amplitude. This is described in more detail in supplementary sections S3 and S4. However, the key aspect is that for each potential fit we are able to additionally simulate the probability that the $\ket{S_{2,8}}$ transitions into $\ket{T_-}$ during the 4ns ramp in. By combining the two fitting approaches we are able to identify the optimal fit the the hole-spin system. In fact of the $\approx$ 50 possible parameter combinations <10 seemed to show reasonable state occupation probabilities, and one optimal fit was clearly identified. For more details see supplementary section S4 for discussion on rejected fits.

The parameters for the optimal fit are
\begin{equation*}
t_c = 13.7\pm0.1 \mu\text{eV} \\
\end{equation*}
\begin{equation*}
     \tensor{g}_L = \begin{pmatrix}
          -0.78 & -1.13 & -1.45 \\
          -1.13 & 0.85 & -0.27 \\
          -1.45 & -0.27 & 1.91 \\
    \end{pmatrix}
\end{equation*}

\begin{equation*}
    \tensor{g}_R = \begin{pmatrix}
          -0.96 & -0.94 & -1.47 \\
          -0.94 & 0.78 & -0.74 \\
          -1.47 & -0.74 & 1.82 \\
    \end{pmatrix}
\end{equation*}
and
\begin{equation*}
    \vec{t}_{so} = \begin{pmatrix}
          -37\pm2  \\
          107\pm4  \\
          0 \pm 20  \\
    \end{pmatrix} \text{neV}
\end{equation*}
We use labels $\tensor{g}_L$ and $\tensor{g}_R$ to indicate a `left' and `right' g-tensor respectively, however this experiment is unable to assign $\tensor{g}_L$ or $\tensor{g}_R$ specifically to the P1 or P2 dots. Uncertainty in each of the g-factors was on the order of $\pm$ 0.02. The optimal fit here provides the parameters used for the solid lines in Figure \ref{fig:Figure2} of the main text. Figure \ref{fig:extfigure-5} a) shows an illustration of the optimal fit g-tensors and spin orbit vector $\vec{t}_{so}$. Note that the principle axes of the left and right g-tensor are slightly misaligned with respect to each other. 

\subsection*{Data Availability}
Data and code are available upon request. 

\subsection*{ACKNOWLEDGMENTS}
This work was funded by the Australian Research Council (Grants No. DP200100147, and No. FL190100167) and the U.S. Army Research Office (Grant No. W911NF-23-1-0092). A.R.H acknowledges an ARC industrial laureate fellowship (IL230100072). R.S.E. acknowledges the SNSF NCCR SPIN International Mobility Grant. Devices were made at the New South Wales node of the Australian National Fabrication Facility. All authors thank A. Saraiva, A. Sarkar, N. Dumoulin Stuyck and E. Vahapoglu for valuable discussions. 

\subsection*{Author Contributions}
S.D.L performed the experiments and analysis. F.E.H and W.H.L fabricated the device. Z.W and S.D.L developed the model for $H_{ST}$. D.J.H and S.D.L developed the QuTip code used to simulate hole spin-dynamics. J.H.H and C.C.E produced the 3D model Fig. \ref{fig:Figure1}a). R.S.E and S.D.L performed fitting of $H_{ST}$. S.D.L wrote the manuscript with input from all co-authors. All authors contributed to discussion and planning. A.R.H supervised the project.

\end{multicols}
\small
\section*{Extended Data Figures}

\renewcommand{\thefigure}{E\arabic{extfigure}} 

\begin{figure*}[t]
        \stepcounter{extfigure} 
	\centering
	\includegraphics[width=16cm]{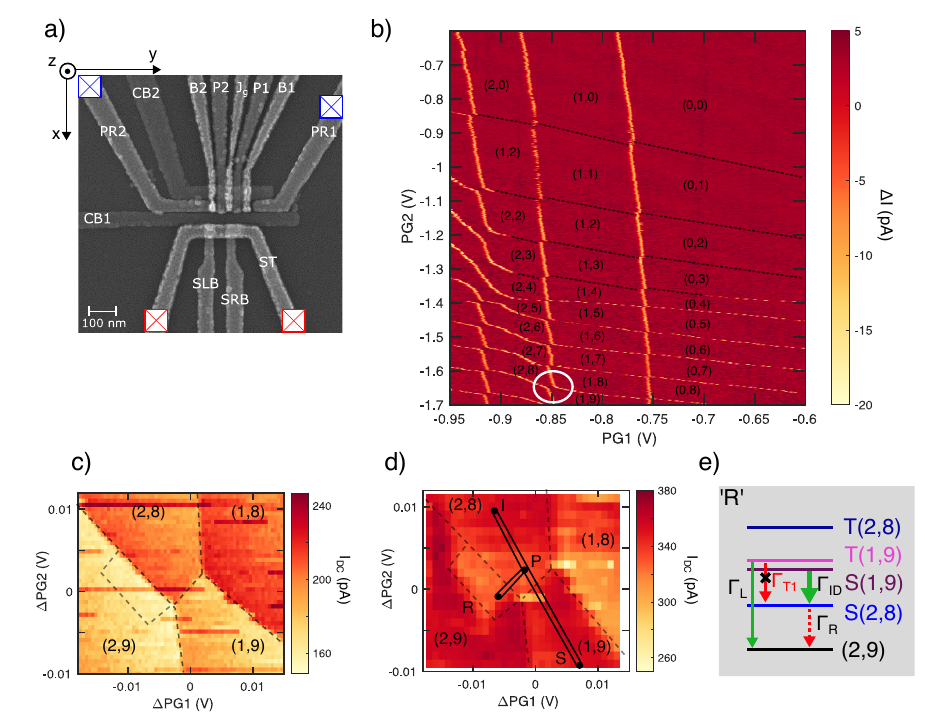}
	\caption[]{\small \textbf{Device and stability diagram}. a) SEM image of a nominally identical device that used here. The bottom electrodes (ST, SLB and SRB) are used to form a SET charge sensor, with two n-type ohmics indicated in red. The top electrodes (PR1, B1,  P1, $J_g$, P2, B2 and PR2) are used to form a double quantum dot, as indicated in Figure\ref{fig:Figure1}a) of the main text. Gates CB1 and CB2 are used to provide x-axis confinement to the double dot. The hole double quantum dot is connected to two p-type ohmic contacts, indicated by the blue squares. b) Full stability diagram of the device down to the last hole. The colour axis is proportional to $dI/dV_{P1}$, where I is the current through the SET charge sensor. The (2,8)-(1,9) operating point is indicated by the white dashed circle. c) A zoom in of the (2,8)-(1,9) charge stability diagram. The colour scale is the DC sensor current in pA. d) DC sensor current around the (2,8)-(1,9) transition when applying a I-S-P-R (I-S-R) pulse sequence and sweeping the DC voltage on the plunger gates P1 and P2. After initialising at I, pulsing to the S location produces a mix of $\ket{S_{1,9}}$ and $\ket{T_0}$. The dwell time at S was chosen to maximise the $\ket{T}$ probability. An enhancement region at the (2,8)-(2,9) transitions allows latched Pauli-spin-blocked readout. This enhancement region is removed when the pulse is reversed (ie M-P-S-R). e) Diagram of the meta-stable states and the expected transitions at the readout position R. the $\ket{T_0}$ state is Pauli-spin-blocked from transiting to $\ket{S_{2,8}}$, but can tunnel to the (2,9) ground state at a rate defined by the left dot tunnel rate ($\Gamma_L>10$~MHz). $\ket{S_{1,9}}$ will quickly transition to the $\ket{S_{2,8}}$ due to the rapid interdot tunnel rate ($t_c>1$~GHz), and will then transition to the (2,9) ground state at a rate defined by the right dot tunnel rate ($\Gamma_R<30$~kHz). We used a 20~$\mu$s dwell at the readout point R such that the difference in $\Gamma_L$ and $\Gamma_R$ allowed readout of $\ket{S_{1,9}}$ and $\ket{T_0}$.
	}
	\label{fig:extfigure-1}
\end{figure*} 

\begin{figure*}[p]
	\centering
        \stepcounter{extfigure} 
	\includegraphics[width=15cm]{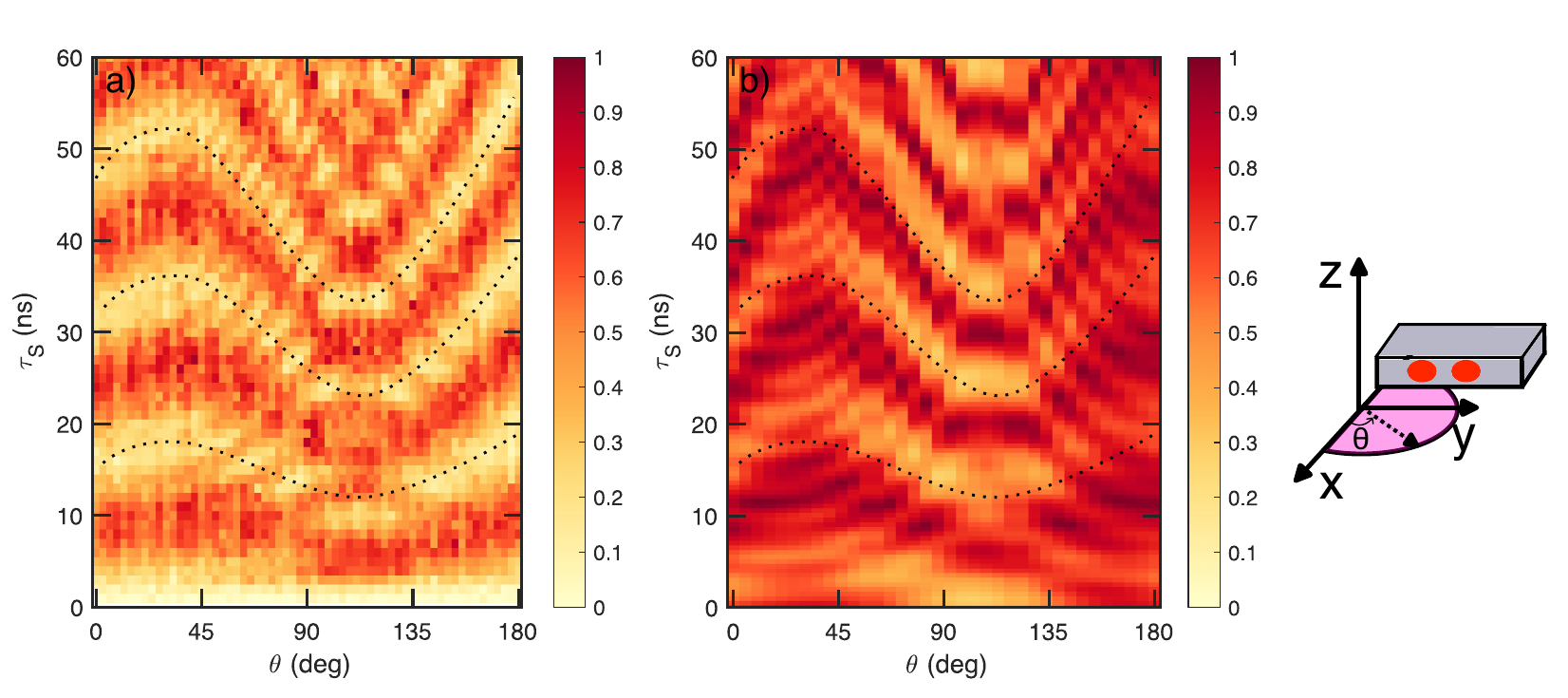}
	\caption[]{\small \textbf{Zoom in of Figure 2 and comparison with QuTip Simulation}. a) The experimental data from Figure \ref{fig:Figure2}c) is reproduced over a shortened time scale. The colour scale is the normalised singlet probability. Black dashed lines serve as a guide to the eye for the anisotropy in the period of the $\Delta$g-oscillations. At $\theta$=0 the slower $\ket{S}$$\leftrightarrow$$\ket{T_0}$ oscillations are the main component picked up in the FFT data of Figure \ref{fig:Figure2}d). However as $\theta$ increases additional spectral components become more pronounced, with most significant effects observed at $\theta$ = 120$^\circ$. These additional spectral components arise from both $\ket{S}$$\leftrightarrow$$\ket{T_\pm}$ oscillations and $\ket{T_0}$$\leftrightarrow$$\ket{T_\pm}$, which both become more prevalent as $\Delta_{ST}$ becomes both more open and moves to more positive detuning (See Figure \ref{fig:Figure2}e). d) Shows the full simulation of the experimental procedure using the Qutip code and model described in Supplementary sections S1, S3 and S4. The black dashed lines are transposed from a) to show that the simulation captures the key experimentally observed features. We note that the experiment in a) takes approximately 3hrs to perform and is limited by the integration time of the SRS830 lock-in (0.3ms). The simulation in b), which uses QuTip `\textit{sesolve}' function, takes approximately 10 hrs to run, and is limited by the number of PC cores running simultaneously.
	}
	\label{fig:extfigure-Model}
\end{figure*} 

\begin{figure*}[p]
	\centering
        \stepcounter{extfigure} 
	\includegraphics[width=16cm]{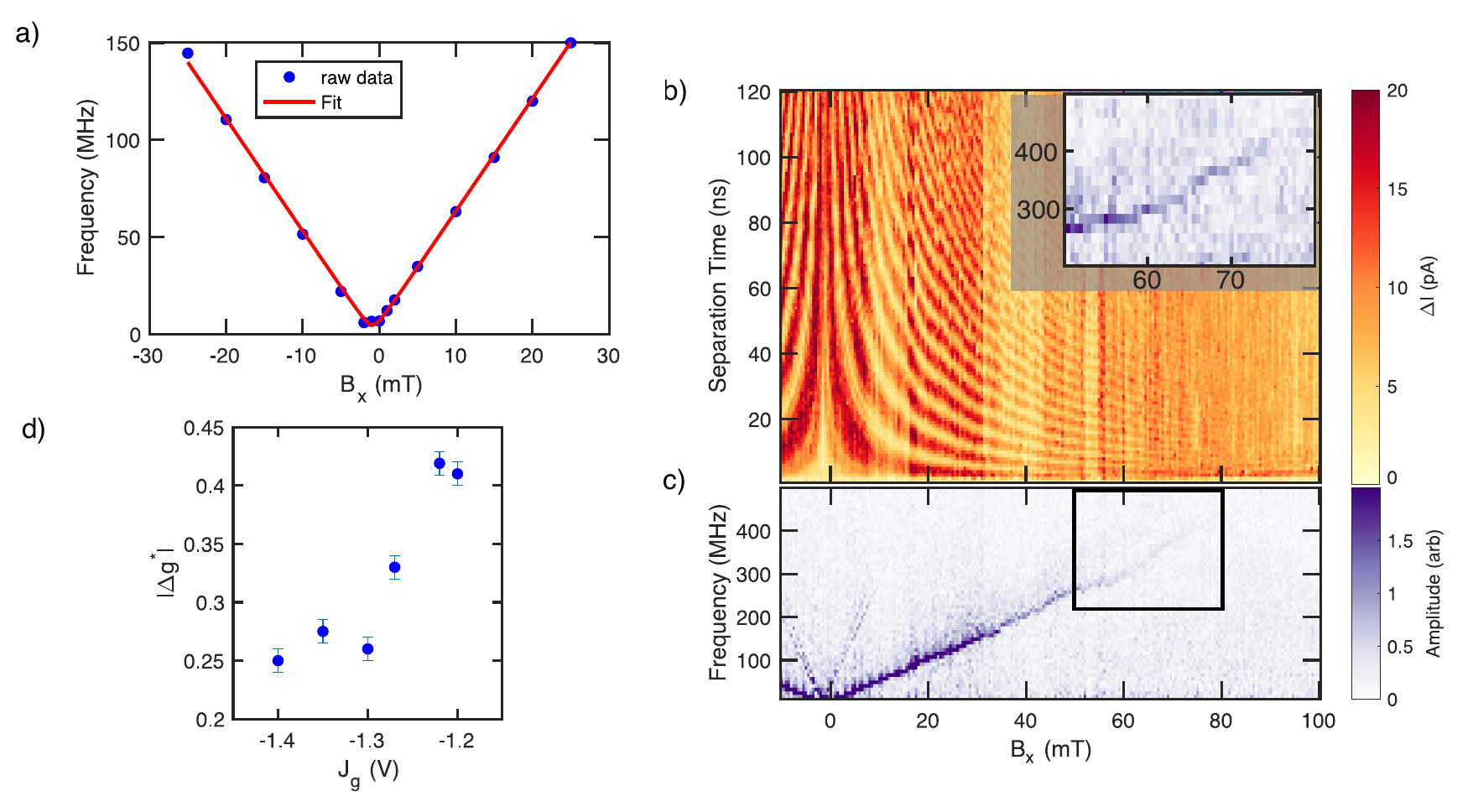}
	\caption[]{\small \textbf{Additional analysis of singlet-triplet oscillations}. a) The singlet-triplet oscillation frequency is plotted as a function of the in-plane magnetic field $B_x$. The solid red line shows the best fit of the raw data to Eqn. \ref{eqn:Est}. From the fit we extract the |$\Delta g*$| for this field orientation, and the residual exchange coupling $J$. b) Shows singlet-triplet oscillations for an in-plane magnetic field up to $B_x$=100 mT. c) The FFT at each |$B_x$| from b) shows the frequency of the singlet-triplet oscillations. A zoom-in of the black rectangle are is shown as in the inset of b), revealing that singlet-triplet oscillations persist to 400~MHz. d) Shows the extracted $\Delta g$ as a function of the $J_g$ gate voltage, demonstrating in-situ control of $\Delta g$. 
	}
	\label{fig:extfigure-2}
\end{figure*}

\begin{figure*}[p]
	\centering
        \stepcounter{extfigure} 
	\includegraphics[width=12cm]{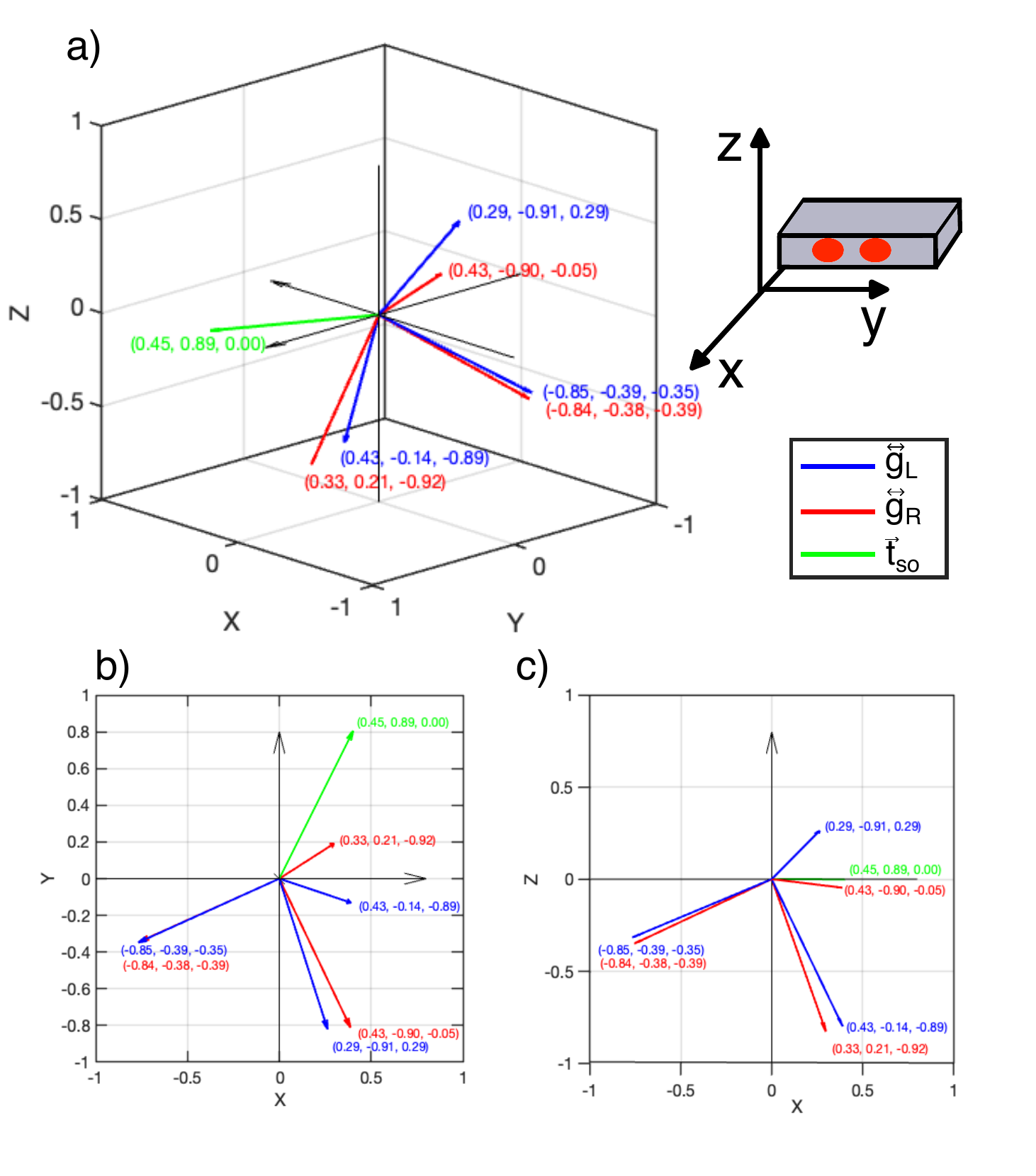}
	\caption[]{\small \textbf{Fitting parameters for the g-tensors and spin orbit vector}. a) The principle axes of the optimal $\tensor{g}_L$ (blue) and $\tensor{g}_R$ (red) are indicated as normalised vectors in the x-y-z plane. The full g-tensors are presented in the methods section. There is a clear misalignment between the principle axes of the left and right g-tensors. The spin-orbit vector is plotted in green. Normalised vectors are used to compare orientations of each g-tensor and the spin-orbit vector, and the components of each normalised vector are indicated. b) and c) show 2D projections of a) in the x-y plane and the x-z plane respectively. 
	}
	\label{fig:extfigure-5}
\end{figure*}

\end{document}


\begin{center}
		
		{\large{\bf
			Supplementary Material: A singlet-triplet hole-spin qubit in MOS silicon.\\
                
			}}
			
			
			\vskip0.5\baselineskip
			
			{\bf
				S. D. Liles$^{1,\dagger}$, D. J. Halverson$^{1}$, Z. Wang$^{1}$, A. Shamim$^{1}$, R. S. Eggli$^{2}$, I. K. Jin$^{1,3}$, J. Hillier$^{1}$, K. Kumar$^{1}$, I. Vorreiter$^{1}$, M. Rendell$^{1}$, J. H. Huang$^{4,5}$, C. C. Escott$^{4,5}$, F. E. Hudson$^{4,5}$, W. H. Lim$^{4,5}$, D. Culcer$^{1}$, A. S. Dzurak$^{4,5}$, A. R. Hamilton$^{1}$  
			}
			
			\vskip0.5\baselineskip
			
			{\it
				$^{1}$School of Physics, University of New South Wales, Sydney NSW 2052, Australia.\\
				$^{2}$Department of Physics, University of Basel, Klingelbergstrasse 82, CH-4056 Basel, Switzerland.\\
				$^{3}$ RIKEN, 2-1, Hirosawa, Wako-shi, Saitama 351-0198, Japan.\\
				$^{4}$School of Electrical Engineering and Telecommunications,\\
				University of New South Wales, Sydney NSW 2052, Australia.\\ 
    			$^{5}$Diraq, Sydney NSW, Australia.\\ 
			}
			
			%
			
			\let\thefootnote\relax\footnote{$\dagger$ Corresponding author - s.liles@unsw.edu.au\\}
\\

\end{center}	
\renewcommand{\thesection}{S\arabic{section}}
\renewcommand{\theequation}{S\arabic{equation}}
\renewcommand{\thefigure}{S\arabic{figure}}

\section{5x5 Singlet-Triplet Hamiltonian}
Here we present full details of the 5x5 Hamiltonian $H_{ST}$, which is used to model the singlet triplet system. We include the Zeeman Hamiltonian in the form 
\begin{equation}
    H_z = \frac{\mu_B}{2} ( (\tensor{g}_L \cdot \Vec{B}) \cdot \Vec{\sigma}_L) \otimes \mathbb{1}_R + \mathbb{1}_L \otimes ((\tensor{g}_R \cdot \Vec{B})\cdot \Vec{\sigma}_R) )
    \label{eqn:Hz}
\end{equation}
where $\tensor{g}_{L,R}$ are the 3x3 g-tensors for the left and right dot, $\mu_B$ is the Bohr magneton, $\vec{B}$ is the magnetic field with respect to the lab (x,y,z) frame, and $\vec{\sigma}_{L,R}$ are the pauli vectors for the left and right dot. We make no prior assumptions that $\tensor{g}_L$ or $\tensor{g}_R$ are diagonalizable in the same eigenbasis, allowing each g-tensor to exhibit independent anisotropy. This is equivalent to allowing the principle axes of each g-tensor to be independent. As a result, and for convenient comparison with the experiment, we define $\Vec{B}$ in the lab frame of reference. 

Using these conventions the g-tensor in the lab frame-of-reference is defined as, 
\begin{equation}
    \tensor{g}_i =\begin{pmatrix}
          g^i_{xx} & g^i_{xy} & g^i_{xz} \\
          g^i_{xy} & g^i_{yy} & g^i_{yz} \\
          g^i_{xz} & g^i_{yz} & g^i_{zz} \\
    \end{pmatrix} 
\end{equation}
where $i$ indicates $L$ or $R$ for the left and right dot. The magnetic field is defined as $\Vec{B} = (B_x,B_y,B_z)$, where each component is the magnetic field in the lab x,y, and z-axes respectively (see Figure 1a of main text for orientation).  Consequently, the term in the Zeeman Hamiltonian $(\tensor{g}_i \cdot \Vec{B})$ is evaluated as,
\begin{equation}
    \tensor{g}_i \cdot \vec{B} =\begin{pmatrix}
          g^i_{xx} B_x + g^i_{xy} B_y + g^i_{xz} B_z \\
          g^i_{xy} B_x + g^i_{yy} B_y + g^i_{yz} B_z  \\
          g^i_{xz} B_x + g^i_{yz} B_y + g^i_{zz} B_z  \\
    \end{pmatrix} .
\end{equation}

In the main text, we define the difference in the Zeeman energy between the left and right dot as
\begin{equation}
    \Delta E_z = |\Delta g^*| \mu_B |\vec{B}|
\end{equation}
where $\Delta g^* = g^*_L - g^*_R $, and $g^*_{L,R}$ are the effective g-factors of the left and right dot respectively. The effective g-factor is the g-factor the is observed for a specific magnetic field orientation and can be evaluated from the respective g-tensor as,
\begin{equation}
    g^*_i = \frac{|\tensor{g_i} \cdot \vec{B}|}{|\vec{B}|}.
\end{equation}

We include a spin-orbit Hamiltonian of the form,
\begin{equation}
    H_{so} = (\Vec{t}_{so}\cdot \Vec{\sigma}_L)\otimes \mathbb{1}_R  + \mathbb{1}_L \otimes ( \Vec{t}_{so} \cdot \Vec{\sigma}_R)
   \label{eqn:Hso}
\end{equation}
where the spin-orbit vector is defined by components in the lab frame such that $\Vec{t}_{so} = (t_x,t_y,t_z)$.

Finally, we include the orbital Hamiltonian in the standard form,
\begin{equation}
    H_{orb} = \epsilon\ket{S_{2,8}}\bra{S_{2,8}} + \sqrt{2}t_c ( \ket{S}\bra{S_{2,8}} + \ket{S_{2,8}}\bra{S})
    \label{eqn:Horb}
\end{equation}
where $\epsilon$ is the detuning energy and $t_c$ is the tunnel coupling between the left and right dot.

The full 5x5 model for the two-hole singet-triplet system is then given by
\begin{equation}
    H_{ST} = H_z + H_{so} + H_{orb}.
    \label{eqn:HST}
\end{equation}

\section{Extended characterisation measurements}
\subsection{Temperature effects}
\begin{figure*}[p]
	\centering
        \stepcounter{extfigure} 
	\includegraphics[width=15cm]{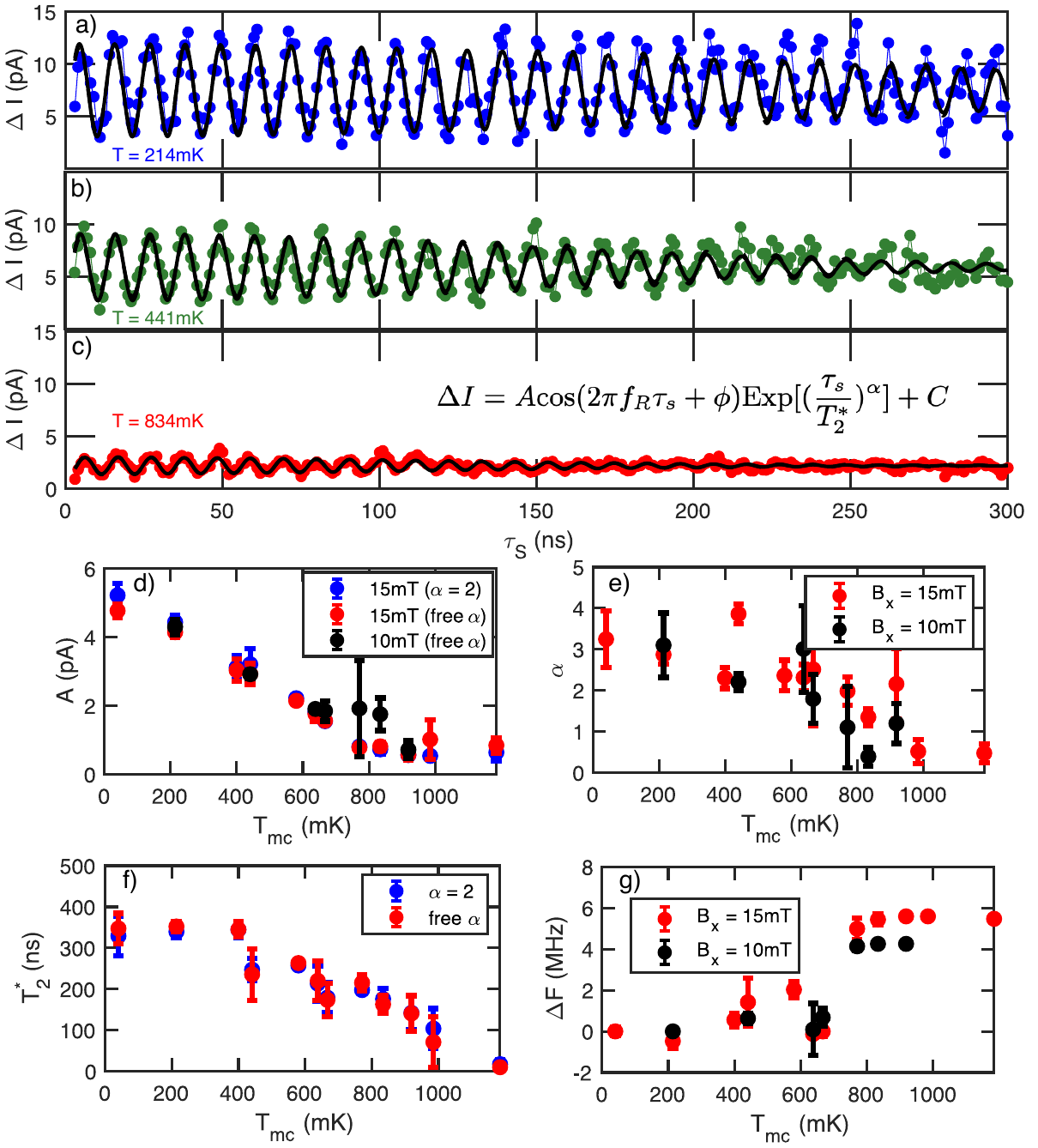}
	\caption[]{\small \textbf{Additional analysis of singlet-triplet oscillation fitting parameters}. a-c) Charge sensor signal as a function of separation dwell time ($\tau_S$) when using the pulse sequence described in Figure 3 of the main text. The measurement is presented at three different mixing chamber temperatures, 214 mK (blue), 441 mK(green), and 834 mK (red). The oscillations in $\Delta I$ are $\Delta g$-driven singlet-triplet oscillations. The solid black line is a best fit of the data to the equation shown in the inset of c). For each data sets, we present two fits. The solid line shows the best fit where $\alpha = 2$. Additionally, a second fit is shown in black dashed line, which is the fit where $\alpha$ is left as a free parameter. However, both fits ($\alpha$ = 2 and $\alpha$ free) are so similar that it is difficult to see the dashed black line. For each measurement the full data set extends to 500 ns. d) We plot the value of \textit{A} extracted for a best fit to oscillations at a range of different mixing chamber temperatures. We have extracted \textit{A} for fits where $\alpha$ = 2 and where $\alpha$ was allowed to be free. Overlap between the two data sets indicates that the value of $\alpha$ does not influence the fit parameter \textit{A}. e) Best fit value of $\alpha$ extracted from oscillations observed at a range of different mixing chamber temperatures. A trend of $\alpha\rightarrow 1$ can be observed as temperature increases. f) Comparison of the best fit coherence time for free and fixed $\alpha$. This demonstrates that the choice of $\alpha$ does not influence the best fit $T^*_2$. g) The shift in oscillation frequency with respect to the base temperature measurement ($f_R$($T_{mc}$= 30 mK) = 30 MHz). 
	}
	\label{fig:extfigure-3}
\end{figure*} 

In Figure \ref{fig:extfigure-3}a-c) we show singlet-triplet oscillations measured at 214 mK, 441 mK and 834 mK respectively. The measurements shown are for $B_x$ = 15 mT and data up to 300 ns is presented, however all data was measured up to 500 ns for accurate extraction of $T_2^*$. In each case the data was fit to the equation shown in the inset of c). This equation is equivalent to Eqn. 4 of the main text, with the condition that $\Delta I$ has been converted to $P_S$ based on the normalisation described above. Figures \ref{fig:extfigure-3}d-f) show the extracted fitting parameters for data measured at a range of different mixing chamber temperatures. We note that the maximum amplitude `A' decreases steadily with $T_{MC}$, starting from base temperature, and dropping to a minimum around 800 mK. This is distinctly different to the $T_2^*$ data, which remains constant up to around 400 mK. We suspect the decrease in `A' is related to a suppression of the visibility of the latched readout. We suggest this may occur as thermal effects equalise the left and right tunneling rates. 

Figure \ref{fig:extfigure-3}e) shows the fit parameter $\alpha$, which is related to the spectral colour of the noise. If the noise power follows a power law, then the noise power is given by $S(f) \propto f^{-\beta}$ and the decay envelope for coherent oscillations is given by exp$(-\tau/T_2^*)^{\beta +1}$. We define $\alpha = \beta +1$, such that $\alpha = 2$ ($\beta = 1$) corresponds to the usual 1/f noise spectrum. Figure \ref{fig:extfigure-3}e) shows value of $\alpha$ obtained for a best fit of the experimental data to the equation displayed in the inset. The data in Figure \ref{fig:extfigure-3}e) shows $\alpha$ decreasing towards 1 as the temperature increases. Previous experiments in other systems have noted that $\alpha$ shows temperature dependence, with the noise tending to become `whiter' (ie $\alpha \rightarrow 1$) as the temperature increases\cite{camenzind2022hole}. The exact mechanism for this change in spectral noise is unclear. Figure \ref{fig:extfigure-3}e) shows that $\alpha$ tends towards 1 as the temperature increases. 

For all fitting in the main text we made the assumption that 1/f noise is the dominant noise source, and fixed $\alpha$ = 2 to simplify the fitting. To ensure this assumption of $\alpha = 2$ doesn't influence the fitting parameters, we performed fitting with a free $\alpha$ and a fixed $\alpha = 2$. Figures \ref{fig:extfigure-3}d) and f) show the best fit value of $A$ and `$T_2^*$' extracted using fixed $\alpha$ and free $\alpha$. In both cases the free choice of $\alpha$ is within the error bars of the fitting, indicating that the values of these fit parameters are reasonably robust against changes in $\alpha$. 

Finally, Figure \ref{fig:extfigure-3}g) shows the extracted singlet-triplet oscillation frequency as a function of mixing chamber temperature. We observe a reproducible shift in the oscillation frequency of 6 MHz as the temperature is increased. We are confident this is not a slow drift in the oscillation frequency since the data were collected in semi-random order. 

\subsection{Full 2$\pi$ characterisation}
\begin{figure*}[ht]
	\centering
        \stepcounter{extfigure} 
	\includegraphics[width=12cm]{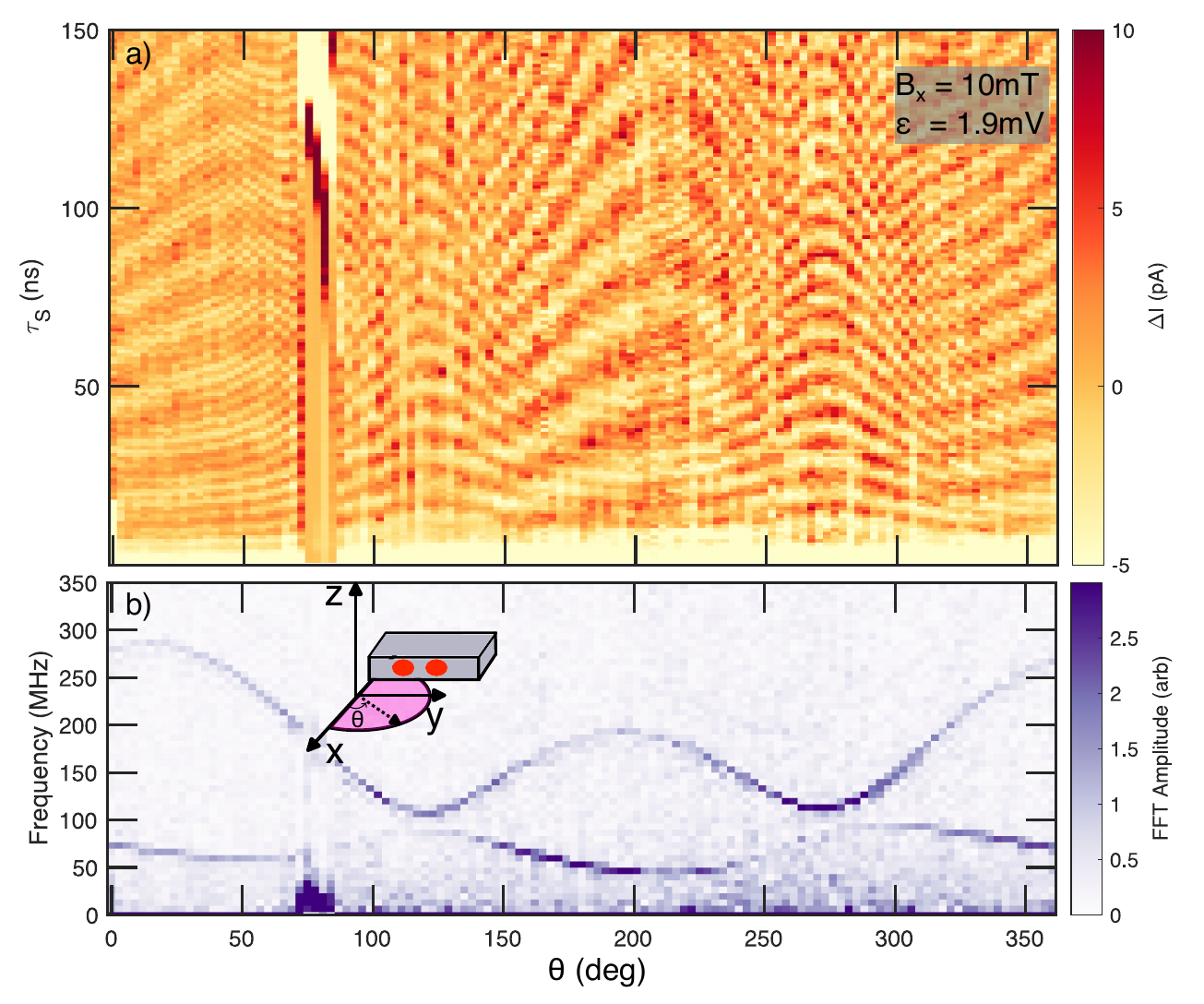}
	\caption[]{\small \textbf{2$\pi$ characterisation of in-plane eigenstate anisotropy}. a) The measurement procedure is the same as that using in Figure 2b) of the main text (I-S-R), however here the experiment is performed over a full 2$\pi$ rotation in-plane. The y-axis shows the separation time and the x-axis indicates the orientation of the magnetic field. Colour scale shows the sensor lock-in signal ($\Delta I$) where several sets of oscillations can be extracted. A shift is the sensor occurred between $\theta$ = 65$^\circ$ and 80$^\circ$, resulting in the spurious signal over this range. b) Shows a FFT of the $\Delta I$ at each magnetic field orientation.
	}
	\label{fig:extfigure-4}
\end{figure*}

Figure \ref{fig:extfigure-4}a) shows the measurement of Figure 2c) extended to a full 2$\pi$ rotation in-plane. Clear anisotropic oscillations can be observed in the sensor signal. For $\tau_S$ less than 4 ns no clear signal is observed. This is due to the effective rise time of the separation pulse. Figure \ref{fig:extfigure-4}b) shows the FFT of the measurement, demonstrating the anisotropy of the eigenstates. These data were obtained using the same pulse sequence used in Figure 2c), which involves an initialisation in S(2,8), a separation pulse to $\epsilon = 1.9$ meV, a dwell at the separation point for time $\tau_S$, and a final pulse to the readout position in order to map the resulting singlet or triplet spin state to the corresponding (2,8) or (2,9) latched charge states.

\section{Model Hamiltonian and simulation of spin dynamics}
The Hamiltonian for the hole double quantum dot is defined in Eqn. \ref{eqn:HST}. This Hamiltonian is defined in the basis of ($\ket{T_+},\ket{T_0},\ket{T_-},\ket{S},\ket{S_{(2,8)}}$). The model Hamiltonian in used in two ways to interpret the experimental data. Firstly, by solving the eigenstates it is possible to determine the expected energy splittings between each level. The predicted level splitting can be compared with the observed oscillation frequency (ie in Figure 2a-d) in order to determine a best fit for the input parameters. In addition, we have implemented a simulation protocol which models the dynamics of a hole-spin following the experimental pulse sequence and including the full Hamilton. This simulation protocol is used to evaluate the probability of loading the leakage states discussed in Figure 2f).

The key parameters for the Hamiltonian are $\tensor{g}_L$, $\tensor{g}_R$, $\vec{t}_{so} = (t_x,t_y,t_z)$, and $t_c$. We consider arbitrary symmetric g-tensors, hence overall this produces 16 free parameters in the model for $H_{ST}$ (Eqn. \ref{eqn:HST}). As shown in Figure 2a-d) of the main text it is possible to observe three independent oscillation frequencies. The best fit procedure uses the anisotropy of the three level splittings to obtain a best fit to the 16 free parameters. We find that $H_{ST}$ (Eqn. \ref{eqn:HST}) provides an excellent fit to the experimental data. However, as a result of the large number of free parameters, when fitting the observed frequencies in Figure 2 to $H_{ST}$, we are unable to obtain a unique fit. Indeed, we determine that there is a wide range of possible g-tensors and spin-orbit combinations, which provide a good fit to the data. 

In order to obtain an optimal fit to the data we consider both the 3 observed FFT frequencies (to obtain best fit parameters), and the FFT amplitude (to assess if the fit parameters support the observed spin-dynamics). As discussed in the main text the FFT amplitude of higher frequency oscillations (ie $\ket{T-}$-$\ket{S}$ oscillations) is dependent on the probability of accessing the $\ket{T-}$ state during the ramp in. Therefore, we also compare the experimental FFT amplitudes with the simulated probability of loading the $\ket{T-}$ state.

We model the spin dynamics of the experiment using the python QuTIP package\cite{johansson2012qutip}. Typical simulation protocol starts by defining the initial state, $\Psi(t = 0)$ as the lowest energy state at large negative detuning ($\Psi (t=O)$ = $\ket{S_{2,8}}$). We simulate a linear ramp from the initial detuning to a positive detuning value, a hold at the positive detuning and a final ramp back to initial detuing. At each increment during this procedure we project the time evolved $\Psi(t)$ back onto the basis states ($\ket{T_+},\ket{T_0},\ket{T_-},\ket{S},\ket{S_{(2,8)}}$) to analyse the evolution of the state.

\begin{figure*}[h]
	\centering
        \stepcounter{extfigure} 
	\includegraphics[width=15cm]{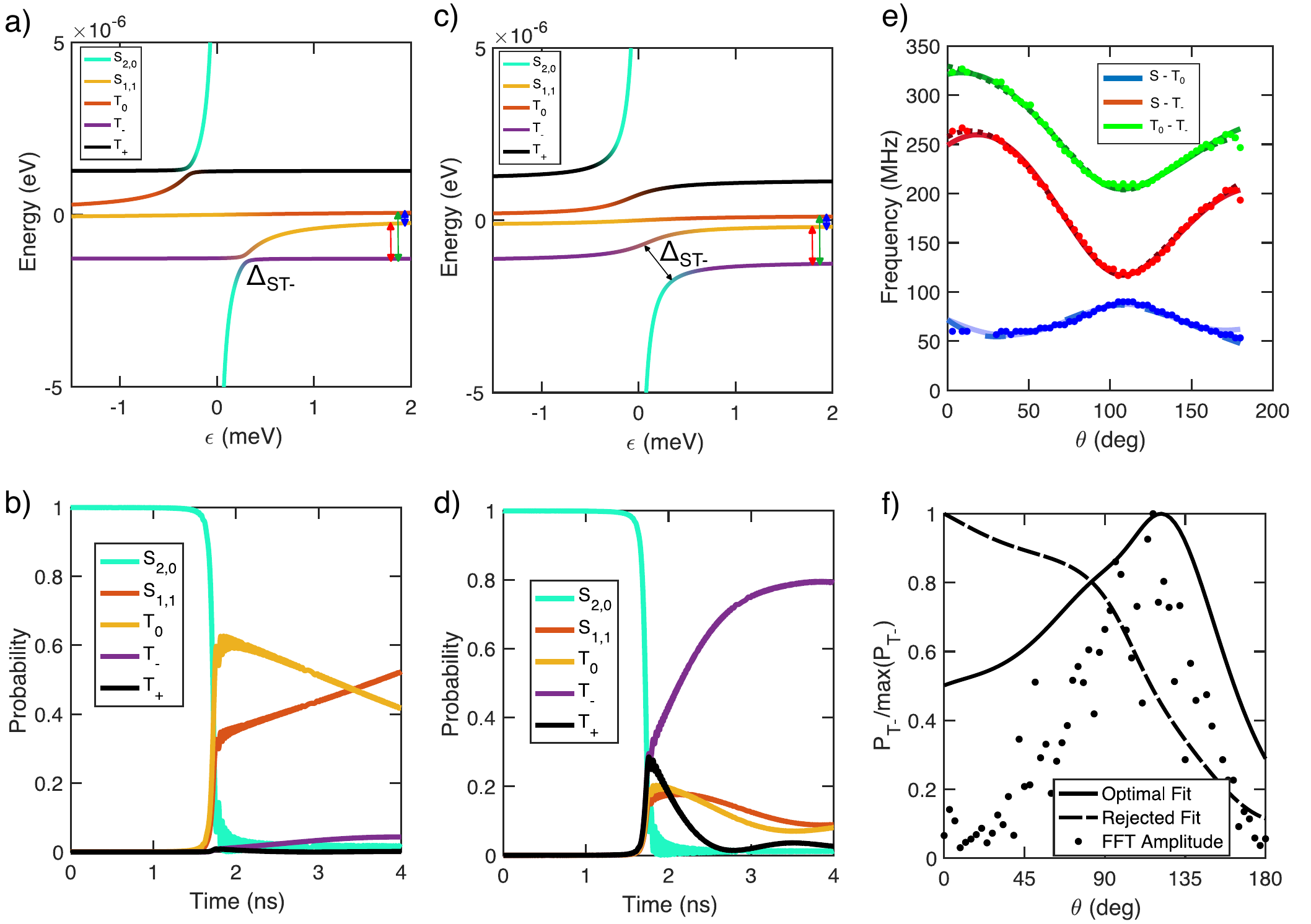}
	\caption[]{\small \textbf{Modelling the hole-spin dynamics}. a) The eigenenergies are plotted as a function of detuning ($\epsilon$) for the `optimum' fit parameters (see Table \ref{tab:FitCompare}). b) Probability of occupation of each basis state during a pulse from $\epsilon$=-1.5 meV to $\epsilon$=+1.9 meV. The simulation begins with the system initialised in $\ket{S_{2,0}}$, which transitions into $\ket{S}$ and $\ket{T_0}$ upon crossing $\epsilon = 0$ at 1.8ns. Due to the $\Delta_{ST_-}$ avoided crossing there is a small residual probability of populating the $\ket{T_-}$ leakage state. c) The eigenenergies are plotted as a function of detuning ($\epsilon$) for the `rejected' fit parameters (see Table \ref{tab:FitCompare}). Here the $\Delta_{ST_-}$ is significantly enhanced due to the large $\Delta g$ present for the rejected fit parameters. Despite the difference in the $\Delta_{ST_-}$ and the overall parameters, the relative level splittings at $\epsilon = 1.9 meV$ are identical between the optimal and rejected fit. This demonstrates the challenge in determining a unique fit using the level splittings alone. d) Probability of occupation of each basis state during a pulse from $\epsilon$ = -1.5 meV to $\epsilon$ = +1.9 meV. The simulation begins with the system initialised in $\ket{S_{2,0}}$, however for the rejected fit parameters the $\ket{S_{2,0}}$ primarily transitions into the $\ket{T_-}$ when crossing $\epsilon = 0$. e) Circles show the level splitting (in frequency) extracted from the experimental data in FFT data in Figure 2d) of the main text. Solid and dashed lines show the level splitting trend for the `optimal' (solid) and `rejected' (dashed) fit parameters. Both fitting parameter sets reproduce the experimental level splitting well. f) Calculated probability of $\ket{S_{2,0}}$ transitioning to $\ket{T_-}$ during the 4 ns separation pulse. The black circles show the amplitude of the experimental $\ket{S}$$\leftrightarrow$$\ket{T_-}$ oscillations. For the rejected fit parameters the trend in $P_{T_-}$ is distinctly different to the experimental data, while for the optimal fit the trend in $P_{T_-}$ co responds well with the experimental trend.  
	}
	\label{fig:extfigure-5}
\end{figure*}

Figure \ref{fig:extfigure-5}a) reproduces the eigenstates of the optimal fit parameters determined in Figure 2 of the main text. To simulate the spin dynamics we model a 4 ns linear ramp in detuning from $\epsilon = $-2 meV to $\epsilon = $1.9 meV (performed for B$_x$=10 mT). Figure \ref{fig:extfigure-5}b) shows the projection of $\Psi(t)$ onto basis states throughout the ramp. Initially $\Psi(t)$ is primarily $\ket{S_{(2,8)}}$ as defined by initialisation. However, as the ramp moves to positive detuning, $\Psi(t)$ evolves into a mix of $\ket{S}$ and $\ket{T_0}$, which are the operating states of the singlet-triplet qubit. However, there is some residual probability that the $\Psi(t)$ state evolves into the leakage states. As discussed in the main text, the likely hood of accessing these leakage states is related to the probability of making an adiabatic transition of either the $\Delta_{ST_-}$ avoided crossing (for $\ket{T_+}$ and $\ket{T_-}$) or the $t_c$ induced avoided cross (for $\ket{S_{2,8}}$).

\section{Identifying optimal fits}
The following section provides a discussion on the full fitting procedure used to identify optimal fitting parameters from the experimental data presented in Figure 2 of the main text. The large number of free parameters in $H_{ST}$ allows many parameter configurations that reproduce the observed anisotropy of the eigenenergies with respect to magnetic field. We constrain the number of fits and identify an `optimal' fit by additionally considering the anisotropy of the $\ket{S}$$\leftrightarrow$$\ket{T_-}$ FFT amplitudes. This is justified since the amplitude of $\ket{S}$$\leftrightarrow$$\ket{T_-}$ oscillations is proportional to the probability of occupying $\ket{T_-}$ ($P_{T_-}$), which we can calculate $P_{T_-}$ using $H_{ST}$. In Table \ref{tab:FitCompare} we present the parameters for the `optimal' fit and an example `rejected' fit.

Figure \ref{fig:extfigure-5}a) reproduces the eigenenergies of the `optimal' fit parameters shown in Figure 2 of the main text. Figure \ref{fig:extfigure-5}b) shows the projection of $\Psi(t)$ onto basis states throughout the 4 ns ramp from the initialisation point in (2,8) to the separation point in (1,9). The simulated ramp is linear with initialisation at $\epsilon = -1.5$ meV and separation at $\epsilon = +2$ meV. As the ramp moves to positive detuning, $\Psi(t)$ primarily evolves into a mix of $\ket{S}$ and $\ket{T_0}$, which are the expected operating states of the singlet-triplet qubit. However, there is some small residual probability that the $\Psi(t)$ state evolves into the leakage states. 

Figure \ref{fig:extfigure-5}c) shows the eigenenergies of the `rejected fit' while Figure \ref{fig:extfigure-5}d) shows the result of the ramp-in when using these eigenenergies. We note that for the `rejected' fit the $\Delta_{ST_-}$ avoided crossing is significantly enhanced. This enhancement of $\Delta_{ST_-}$ results from the large $\Delta g$ obtained in the fit parameters. As a result, the ramp-in shown in Figure \ref{fig:extfigure-5}d) results primarily in the loading of the $\ket{T_-}$ state, which is a leakage state for the singlet-triplet qubit. Comparing Figure \ref{fig:extfigure-5}b) and d) demonstrates that although the `optimal' and `rejected' fit have equal energy splitting at the separation point, the different fit parameters result in very different dynamics during the ramp-in.

Figure \ref{fig:extfigure-5}e) shows the predicted anisotropy of the level splittings for the `optimal' (solid) and `rejected' (dashed) fits for a magnetic field rotation through the x-y plane. The three level splittings shown are the $\ket{S}$-$\ket{T_0}$ (blue), $\ket{S}$-$\ket{T_-}$ (red), and $\ket{T_-}$-$\ket{T_0}$ (green). These are calculated at $\epsilon$ = 1.9 meV. Circles of the respective colours show the frequency extracted from the FFT measurements in Figure 2 of the main text. Although the `optimal' and `rejected' firs produce significant difference in the $\epsilon$ dependence, both reproduce the observed response of the eigenergy splitting to magnetic field orientation. This is because the observed energy splitting is sensitive only to the eigenenergies at the separation point ($\epsilon = 1.9$ meV), and is not influence by the dynamics during the ramp-in.

Figure \ref{fig:extfigure-5}f) shows the simulated $\ket{T_-}$ probability after the 4 ns ramp-in a range of magnetic field orientations around x-y plane. We plot the normalised $\ket{T_-}$ probability, which we define as $P_{T_-}(\theta) / \text{max}( P_{T_-}(\theta))$. The normalised $\ket{T_-}$ probability shows a distinct difference in the anisotropy between the `optimal' and `rejected' fits. For comparison, we assume that the amplitude of the experimentally observed $\ket{T_-}$-$\ket{S}$  oscillations is proportional to the probability of loading into the $\ket{T_-}$. Hence, in Figure \ref{fig:extfigure-5}f) we plot the normalised $\ket{T_-}$-$\ket{S}$  oscillation amplitude (extracted from the amplitude of the FFT peaks in Figure 2c). For the `optimal' fit the trend in $P_{T_-}$ matches the experimental  amplitude well, with both exhibiting a peak around 120$^\circ$ and dropping off asymmetrically on either side of the peak. However, for the `rejected' fit there is significant discrepancy in the observed anisotropy of $P_{T_-}$ with respect to magnetic orientation. By comparing the anisotropy of the $\ket{T_-}$ probability with the experimental FFT amplitude, we are able to identify an `optimal' fit from over 50 different possible best fits to the data.

\section{Optimal fit and g-tensor.}
\begin{table}[h]
  \centering
  \begin{tabular}{|c|c|c|}
    \hline
    Parameter & Optimal Fit & Rejected Fit \\
    \hline
    $t_c$ & (13.7$\pm0.1)\mu$eV & (14.2$\pm0.1)\mu$eV  \\
    \hline
    $\vec{t}_{so} = (t_x,t_y,t_z)$ & (-36.9$\pm2$, 107$\pm4$, 0 $\pm10$ ) neV & (-234$\pm5$, -295$\pm6$, 0$\pm10$ ) neV \\
    \hline
    $\tensor{g}_L$ &  
    $\begin{pmatrix}
          -0.78 & -1.13 & -1.45 \\
          -1.13 & 0.85 & -0.27 \\
          -1.45 & -0.27 & 1.91 \\
    \end{pmatrix} $
    & 
    $\begin{pmatrix}
          0.09 & 0.38 & 0.93 \\
          0.38 & 0.65 & 0.64 \\
          0.93 & 0.64 & -1.56 \\
    \end{pmatrix} $  
    \\

    \hline
    $\tensor{g}_R$ &
    $\begin{pmatrix}
          -0.96 & -0.94 & -1.47 \\
          -0.94 & 0.78 & -0.74 \\
          -1.47 & -0.74 & 1.82 \\
    \end{pmatrix} $  
    &  
    $\begin{pmatrix}
          1.26 & -0.42 & 1.22 \\
          -0.42 & 0.27 & 1.22 \\
          1.22 & 1.22 & -1.73 \\
    \end{pmatrix} $  
    \\

    \hline
  \end{tabular}
  \caption{Best fit parameters for the Optimal fit and a Rejected fit to the experimentally observed FFT frequencies in Figure 2 of the main text. The best fit procedure allows free fitting to 16 parameters; $t_c$, $t_x$, $t_y$, $t_z$, 6 parameters for $g_L$ and 6 parameters for $g_R$. Uncertainty in each of the g-factors was on the order of $\pm$ 0.02. As discussed, no unique fit to the level splittings was possible, with more than 50 possible combinations of the free parameters giving good fits. The Optimal fit was determined by additionally considering the $\ket{T_-}$ probability following a 4 ns ramp-in. The optimal fit here is the fit used for the solid lines in Figure 2 of the main text. The rejected fit is one example of a possible fit, which was rejected based on the $\ket{T_-}$ probability analysis.}
  \label{tab:FitCompare}
\end{table}

The parameters for the optimal fit and a rejected fit are presented in Table \ref{tab:FitCompare}.

\small{

}